\documentclass{aa}

\usepackage{natbib}
\usepackage{graphicx}
\usepackage{txfonts}   
\usepackage{multirow}
\usepackage{hyperref}
\usepackage{placeins}

\titlerunning{43621}

\begin{document}

\title{Constraints on planetary tidal dissipation from a detailed study of Kepler 91b.}

\author{L. Fellay\inst{1,2}\and C. Pezzotti\inst{2} \and G. Buldgen\inst{2} \and P. Eggenberger\inst{2} \and E. Bolmont\inst{2} }

\institute{STAR Institute, University of Liège, 19C Allée du 6 Août, B$-$4000 Liège, Belgium\and Observatoire de Genève, Université de Genève, Chemin Pegasi 51, CH$-$1290 Sauverny, Suisse }
\date{September,  2022}
\abstract{
With the detection of thousands of exoplanets, characterising their dynamical evolution in detail represents a key step in the understanding of their formation. Studying the dissipation of tides occurring both in the host star and in the planets is of great relevance in order to investigate the distribution of the angular momentum occurring among the objects populating the system and to studying the evolution of the orbital parameters. From a theoretical point of view, the dissipation of tides throughout a body may be studied by relying on the so-called phase or time-lag equilibrium tides model in which the reduced tidal quality factor $Q'_{\rm{p}}$, or equivalently the product between the love number and the time lag ($k_{\rm{2, p}}\Delta t_{\rm{p}}$), describe how efficiently tides are dissipated within the perturbed body. Constraining these factors by looking at the current configuration of the exoplanetary system is extremely challenging, and simulations accounting for the evolution of the system as a whole might help to shed some light on the mechanisms governing this process.}
{We aim to constrain the tidal dissipation factors of hot-Jupiter-like planets by studying the orbital evolution of Kepler-91b.}
{We firstly carried out a detailed asteroseismc characterisation of Kepler-91 and computed a dedicated stellar model using both classical and astereoseismic constraints. We then coupled the evolution of the star to the one of the planets by means of our orbital evolution code and studied the evolution of the system by accounting for tides dissipated both in the planet and in the host star.}
{We found that the maximum value for $k_{\rm{2, p}}\Delta t_{\rm{p}}$ (or equivalently the minimum value for $Q'_{\rm{p}}$) determining the efficiency of equilibrium tides dissipation occurring within Kepler-91b is $0.4\pm0.25$ s ($4.5^{+5.8}_{-1.5} \times 10^5$). We constrained these factors by computing the evolution of the planetary orbit and by reproducing the current properties of the Kepler-91 system.}
{We developed a new method to constrain the tidal dissipation factors using the observed eccentricity of a given planet. Our new approach showed that Kepler 91b has dissipation coefficients compatible with colder Jupiter-like planets. When applied to other targets, our new method could potentially give more precise boundary values to the tidal dissipation factors, and determine whether planetary tides dominate the dissipation during the stellar main sequence.}

\keywords{Planet-Star interactions - Planets and satellites: dynamical evolution and stability - Planets and satellites: gaseous planets -  Planets and satellites: individual: Kepler-91b - Asteroseismology }
\maketitle

\section{Introduction}
The late 20$^{th}$ century saw the birth of the field of exoplanetology with the detection of the first exoplanet around a sun-like star, 51 Pegasi-b \citep{Mayor1995}. The improvement of planetary detection and characterisation techniques in the 21$^{st}$ century has allowed a confirmed detection of $5030$ planets (\cite{ExoplanetArchive} discussed NASA’s Exoplanet Archive, which provides data on these systems). Among all the planets discovered to date, hot, massive, close-in planets called hot Jupiters are in the most extreme environment and sometimes found with high eccentricities (\cite{Grunblatt2018, Grunblatt2022}, in paricular when their periods are longer than five days. These discoveries led to an improvement of the theoretical understanding and description of such systems, which can present different architecture to that of our Solar System. The study of the dynamical evolution and stability of planetary systems is fundamental in order to retrieve the formation of these systems in their variety and diversity, and the role played by tidal forces in this context is crucial. 
\\~\\Three channels of formation are favoured for hot Jupiters: in-situ formation, migration caused by the protoplanetary disc interaction, or high-eccentricity tidal migration \citep{Dawson2018}. The last scenario is particularly interesting from a dynamical point of view as the tidal dissipation plays a key role in determining the evolution of the system. Hot Jupiters are characterised by significantly inflated radii when compared with the one of Jupiter \citep{Bodenheimer2001}. \cite{Guillot2002} suggested that irradiation can slow the Kelvin-Helmholtz contraction of hot Jupiters. Moreover, evidence that re-inflation can occur rapidly in post-main-sequence stars are presented in \cite{Grunblatt2016,Thorngren2021}  and \cite{Grunblatt2022},  while \cite{Thorngren2021} also came to the conclusion that a re-inflation could occur during the host star's main sequence (MS). The drastic change of the planetary structure can have an important impact on its dynamics; in particular, a change in the structure may lead to a change in the efficiency of tidal dissipation within the planet.  Constraining the tidal dissipation efficiency is challenging, and only weak constraints have been derived from dynamical studies of extra-solar planets. In the Solar System, \cite{Lainey2009} constrained the tidal dissipation of Jupiter with an accuracy of about 20$\%,$ while Saturn's tidal dissipation was constrained with an accuracy of $30\%$ \citep{Lainey2012}. A constraint on the tidal dissipation of Hot-Jupiters would be a significant improvement in the modelling of a multi-planetary system as such planets can strongly affect the dynamical evolution of smaller planets.
\\~\\In this study, we investigated the dynamical evolution of close-in planets orbiting red giant branch (RGB) stars, aiming to constrain the tidal dissipation of the single hot Jupiter orbiting the RGB star Kepler-91 (KIC 8219268). The particular interest in the Kepler-91 system stems from the presence of a close-in orbiting hot-Jupiter-like planet, namely Kepler-91b, which orbits its host star at one of the smallest semi-major axes over stellar radius ratio ever seen in any catalogue. This planet is expected to undergo strong tidal interaction and to be engulfed by its host star in about 55 million years \citep{Lillo-Box2014b, Lillo-Box2014}. 
\\~\\Asteroseismology has recently become the most accurate method to precisely characterise and model solar oscillating stars. The advent of long space-based photometric surveys such as CoRoT \citep{CoRoT2009}, \textit{Kepler} \citep{KEPLER2010}, and TESS \citep{TESS2014, TESS2015} gave the material to conduct precise asteroseismic studies on various types of stars and in particular exoplanet host stars. As knowledge of the dynamics of a planetary system is proportional to knowledge concerning its host star,  asteroseismology appears to be an ideal way to characterise stars allowing the precise dynamical study of the planetary systems orbiting it.   Among all observed RGB stars, which are evolved solar stars, are particularly interesting physics laboratories. The extended envelope and contracted core of RGB stars allows mixed modes to propagate inside all the layers, meaning that the very deep rotation profile and structure of the star can be probed. The extended convective zone of RGB stars enhances the interaction between the planet and the star through the tidal force \citep{Villaver2009, Villaver2014, Rao2018}. Moreover, during the stellar RGB phase, the whole planetary evolution is drastically speeded up due to the fast and dramatical changes of the stellar properties \citep{Gamow1939}.
\\~\\The strategy used in this work to study Kepler 91b is based on a detailed modelling of the entire system. First, the star is modelled independently with the Code Liégeois d'Evolution Stellaire \citep[CLES,][]{Scuflaire2008a} and the Liège OScillation Code \citep[LOSC,][]{Scuflaire2008b} taking particular care in reproducing the observed stellar classical constrains and asteroseismic quantities. Then the evolution of the star is coupled to the one of the planets with our orbital evolution code \citep{Privitera2016I, Privitera2016II, Meynet2017, Rao2018} to study the newly implemented evolution of the planet orbital eccentricity. In particular, the stellar equilibrium tides are expected to play a key role in the evolution of the system during the stellar RGB phase \citep{Villaver2009, Villaver2014, Rao2018}.
The physics and the dissipation of the planetary tides are closely linked to dissipation factors: $k_{\rm{2, p}}\Delta t_{\rm{p}}  $ \citep{Darwin1879} or $Q'_{\rm{p}}$ \citep{ Goldreich1963, Goldreich1966}. In our study, we tried to determine boundaries to these quantities by estimating the maximum eccentricity reachable by the planet for different values of $k_{\rm{2, p}}\Delta t_{\rm{p}}$ or $Q'_{\rm{p}}$.
\\~\\As a good description of the structure of the star and its evolution is required to precisely study the dynamical evolution of its planet, we started by carrying out a detailed stellar modelling of Kepler 91, which is presented Sect.~\ref{sec_Stellar_model}. Section~\ref{sec_KEP91b_presentation} is dedicated to a brief presentation of the planetary system and properties. The orbital evolution code used to model the evolution of the planet and the equations describing the evolution of the orbital eccentricity are presented in Sect.~\ref{section_code}. The detailed study of the past dynamical evolution of Kepler 91b and the consequence of our study on the tidal dissipation factors are presented in Sect~\ref{sec_KEP91b_modelling}. Finally, a review of our results and the potential improvements to our method are presented in Sect.~\ref{sec_conclusion}

\section{Stellar models and properties}\label{sec_Stellar_model}
To study the evolution of the planetary system around Kepler-91, a reliable model of the evolution of the star should be obtained. To that end, we combined global and local minimisation techniques using seismic and non-seismic constraints to derive a robust model of our target. The coupling between the star and the planetary system is comprehensively taken into account by the orbital evolution code \citep{Privitera2016I, Privitera2016II, Rao2018}; thus, the star was considered as isolated and modelled independently.

\subsection{Observational constraints}
The star was modelled based on asteroseismic and classical constraints as in \cite{Lillo-Box2014}. In Table~\ref{table_classical_constraints}, we summarise the global seismic indexes as well as the classical constraints for Kepler-91, while a table with the full seismic data adopted is given in Appendix \ref{table_data_sismo}.  

\begin{table}[h]
\centering
\caption{Classical observational constraints obtained by spectroscopy in the top panel. Global seismic parameters are presented in the bottom panel \citep{Lillo-Box2014}.}\label{table_classical_constraints}
\begin{tabular}{ll}
\hline\hline
\multicolumn{2}{l}{\multirow{2}{*}{Observational constraints}} \\
\multicolumn{2}{l}{}                                       \\ \hline
Metallicity [Fe/H]            & $0.11 \pm 0.07$      \\
Effective Temperature {[}$K${]}     & $4550 \pm 75$        \\
Luminosity {[}$L_\odot${]}          & $16.745 \pm  2.578$    \\ \hline
\multicolumn{2}{l}{Large separation $\Delta\nu=9.39 \pm 0.22\ \mu \rm{Hz} $}             \\
\multicolumn{2}{l}{Frequency of max. power $\nu_{max}=108.9 \pm 3.0\ \mu \rm{Hz} $} 
\\ \hline
\end{tabular}
\end{table}
In Table~\ref{table_classical_constraints}, $\nu_{max}$ denotes the frequency of the oscillation spectrum where the higher amplitude is reached.
The luminosity presented in Table \ref{table_classical_constraints} was computed from the following formula \citep{Casagrande2014}:
\begin{equation}
\resizebox{\hsize}{!}{$
\log\left(\frac{L}{L_\odot}\right) = -0.4\left(m_\lambda + BC_\lambda -5\log d + 5 - A_\lambda -M_{\mathrm{bol},\odot}\right),$}
\end{equation}
where $m_\lambda$, $BC_\lambda$, and $A_\lambda$ are the apparent magnitude, bolometric correction, and extinction in a given band $\lambda, $respectively. We used the 2MASS $K$-band magnitude properties. The bolometric correction is estimated using the code written by \citet{Casagrande2014,Casagrande2018}, and the extinction is inferred with the \citet{Green2018} dust map. A value of $M_{\mathrm{bol},\odot} = 4.75$ is adopted for the solar bolometric magnitude \citep{Mamajek2015}. Using Gaia DR2 \citep{Gaia2018}, we obtain a luminosity value of $L=16.745 \pm  2.578\ L_{\odot}$.

\subsection{Stellar modelling}
The stellar modelling procedure follows that of \cite{Fellay2021}. Our modelling is divided in two distinct phases; a first modelling step is carried out with the AIMS software \citep{AIMS2016, AIMS2019} using classical and asteroseismic constraints. The second modelling step is carried out with a Levenberg-Marquardt minimisation technique coupled with the Code Liégeois d'Evolution Stellaire \citep[CLES,][]{Scuflaire2008a} and the Liège OScillation Code \citep[LOSC,][]{Scuflaire2008b},  aimed at reproducing the whole oscillation spectrum of Kepler-91 by computing models on the fly rather than using a pre-defined grid. 
\\~\\The first modelling step is required to determine a first estimate of the stellar properties. It is then supplemented by a deeper local exploration of the parameter space to reproduce the full oscillation spectrum.
We used the AGSS09 \citep{Asplund2009} abundances for both AIMS and the Levenberg-Marquardt method. For both modelling steps, the models use the FreeEOS equation of state \citep{Irwin2012}, the OPAL opacities \citep{Iglesias1996}, the $T(\tau)$ relation from Model-C of \citet{Atmosphere1981} for the atmosphere, the mixing length theory of convection implemented as in \citet[][]{Cox1968}, and the nuclear reaction rates of \cite{Reaction2011}. The constraints and free parameters for the AIMS modelling are summarised in the second column of Table~\ref{table_model_param}, while the fourth column presents the constraints and free parameters of the Levenberg-Marquardt modelling. In Table~\ref{table_model_param}, $L$ denotes the luminosity of the star, [Fe/H] its observed metallicity, $X_0$ the initial hydrogen mass fraction, and $Z_0$ the initial metal mass fraction. Convection is controlled by two parameters, the classical mixing-length parameter $\alpha_{MLT}$ and $\alpha_{over}$ characterising the length of the core overshooting region with the same formalism as the MLT. The temperature gradient in the overshooting region is assumed adiabatic in both modelling steps. The mixing-length parameter was kept at a solar-calibrated value for all tracks of the AIMS grid. While fitting the individual radial modes, we took into account the impact of surface effects by using the two-term surface correction of \cite{Ball2014} in AIMS.

\begin{table}[h]
\centering
\caption{Summary of the constraints and free parameters used for both AIMS and the Levenberg-Marquardt modelling steps.  $a_3$ and $a_{-1}$ denote the surface correction coefficient defined in \cite{Ball2014}. $\mathcal{U}$ denotes the uniform priors used in AIMS.   }\label{table_model_param}
\resizebox{\hsize}{!}{\begin{tabular}{clll}
\hline\hline
\multirow{2}{*}{}                                    & \multirow{2}{*}{AIMS} &\multirow{2}{*}{AIMS's priors}& \multirow{2}{*}{Levenberg} \\
                                                     &                       &      &                      \\ \hline
\multirow{6}{*}{Constraints}                         & [Fe/H]        &         & $L$                   \\
                                                     & $\nu_{max}$        &   &       $T_{\rm{eff}}$        \\
                                                     & $7\ \nu_{\ell=0}$   &           &      [Fe/H]             \\
                                                     & $T_{\rm{eff}}$        &     &         First $\nu_{\ell=0}$           \\
                                                     &               &        &  $7 d_{02}$      \\ \hline
\multicolumn{1}{l}{\multirow{6}{*}{Free parameters}} &  $M$  & $\mathcal{U}[1.00-2.22]\ M_\odot$            & $M$                   \\
\multicolumn{1}{l}{}                                 & Age        &  $\mathcal{U}[0-14]$ Gyr         & Age                        \\
\multicolumn{1}{l}{}                                 & $X_0$    &      $\mathcal{U}[0.68-0.72]$       & $X_0$                      \\
\multicolumn{1}{l}{}                                 & $Z_0$      &   $\mathcal{U}[0.010-0.034]$        & $Z_0$                      \\
\multicolumn{1}{l}{}                                 & $a_3$      &      no priors     & $\alpha_{\rm{MLT}}$             \\
\multicolumn{1}{l}{}                                 & $a_{-1}$     &    no priors        & $\alpha_{\rm{over}}$            \\ 
\hline
\end{tabular}
}
\end{table}

\begin{figure}[h!]
\centering
\includegraphics[width=\hsize]{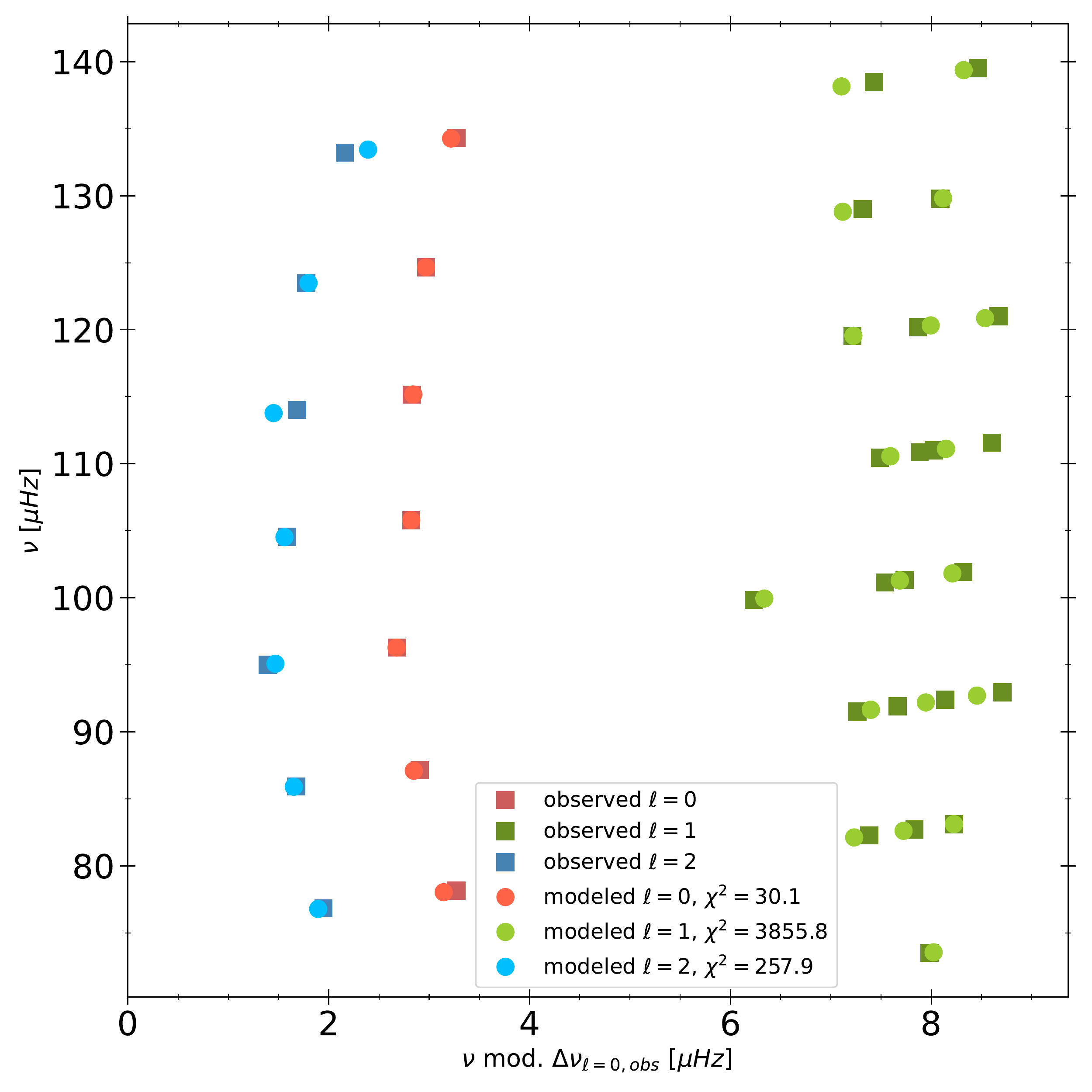}
\caption{Echelle diagram of Kepler 91 comparing observed oscillation frequencies (listed in Table~\ref{table_data_sismo}) represented as squared points and theoretical oscillation frequencies represented as circle points. The whole modelled oscillation spectrum for the $\ell=1$ modes is included with shaded points. The difference between observed and modelled data is quantified, in the legend, as a $\chi^2$ for each spherical order $\ell$. Finally, the error bars on the observed frequencies are included in the data points and not clearly distinguishable here.}
\label{fig._echelle_diagram}
\end{figure}

During the last modelling step, we ensure that our models reproduce the seven asteroseismic quantities $d_{02} $ defined as 
\begin{equation}
d_{02,n}=\nu_{n,0}-\nu_{n-1,2},
\end{equation}
where $\nu_{n,0}$ is the frequency of the oscillation mode with a radial order $n$ and a degree $\ell=0$.
The $d_{02}$ are less affected by the so-called surface effects. The reproduction of the $d_{02}$ coupled with the first radial mode ensures the reproduction of the behaviour of the $\ell=0$ and $\ell=2$ modes; however, due the mixed character of the $\ell=1$ modes we took particular care to verify that the global seismic pattern was reproduced with an Echelle diagram. 

\begin{figure}[h!]
\centering
\includegraphics[width=\hsize]{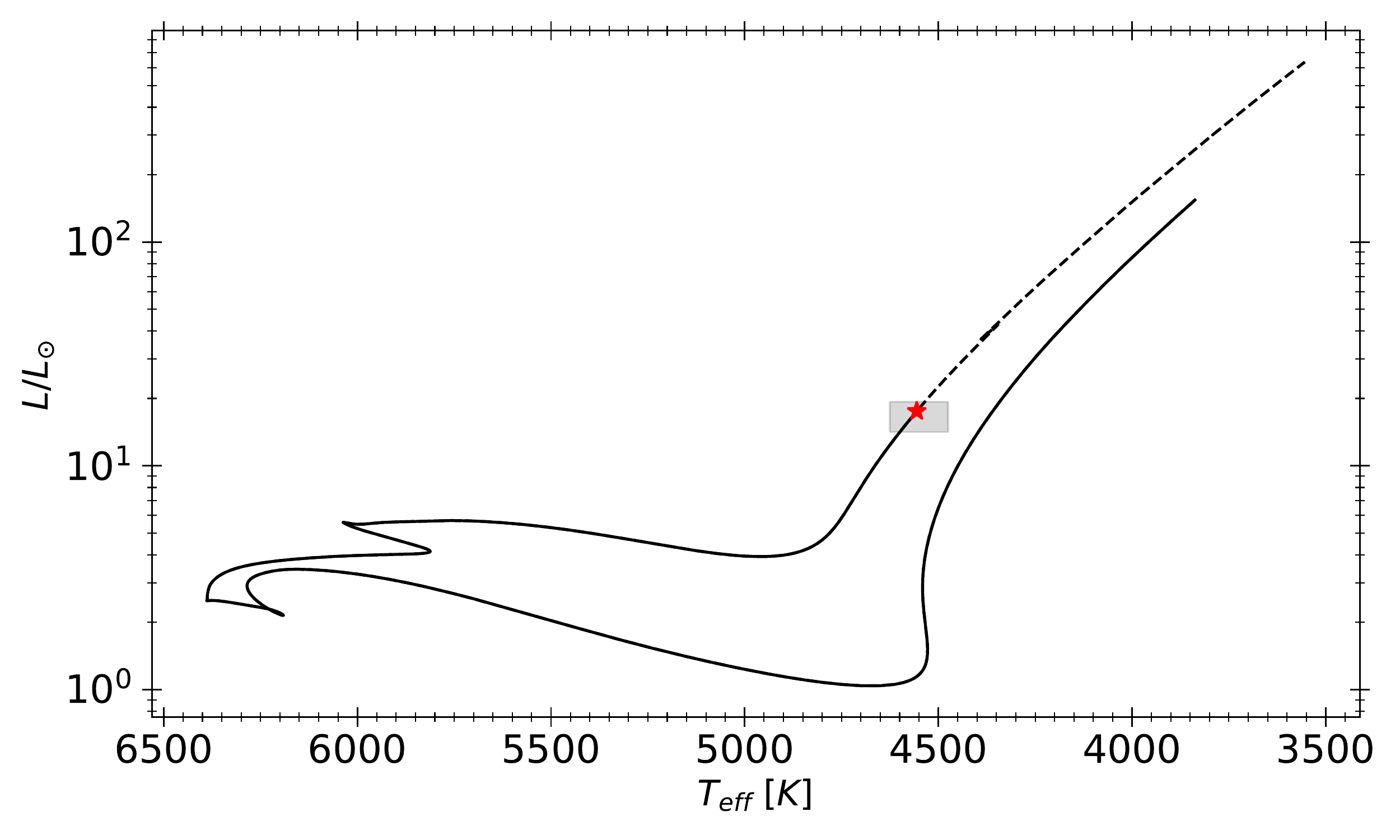}
\caption{Hertzsprung–Russell diagram of the evolution of Kepler 91 from the pre-Main Sequence to the modelled actual state of the star and the its future evolution with dotted line. The grey area corresponds to the constraints on the effective temperature and luminosity listed in Table~\ref{table_classical_constraints}.}
\label{fig._hr_diagram}
\end{figure}

\subsection{Stellar models}

The final solutions of our structural modelling are given in Table~\ref{table_stellar_properties}. The agreement obtained in terms of pulsation frequencies is illustrated by the Echelle diagram shown in Fig.~\ref{fig._echelle_diagram} for the whole pulsation spectrum, while the Hertzsprung–Russell diagram of the star modelled with the Levenberg-Marquardt minimisation technique presented in Fig.~\ref{fig._hr_diagram}. 

\begin{table}
\centering
\caption{Non-seismic stellar properties of Kepler-91 obtained in the different modelling steps.}\label{table_stellar_properties}
\resizebox{\hsize}{!}{\begin{tabular}{llll}
\hline\hline
\multirow{2}{*}{Stellar parameters} & \multirow{2}{*}{AIMS} &  \multirow{2}{*}{Levenberg} \\
                                    &                                                          &                        \\ \hline
Mass {[}$M_\odot${]}                & $1.278 \pm 0.015$                              & $1.307 $           \\
Radius {[}$R_\odot${]}              & $6.24 \pm 0.03$                               & $6.299  $           \\
Mean Density {[}$g/cm^3${]}         & $ 0.00739\pm 0.00001$                            & $ 0.00737 \ $         \\
Metallicity $Z_0/X_0$           & $0.027 \pm 0.02$                             & $0.0234  $          \\
Effective temperature {[}$K${]}     & $4771 \pm 15$                                   & $4578  $            \\
Age {[}$Gyr${]}                     & $4.968 \pm 0.173$                               & $4.8187  $           \\
Luminosity {[}$L_\odot${]}          & $ 18.13 \pm 0.25$                              & $ 15.709  $          \\
$\alpha_{\rm{MLT}}$                      & \textunderscore                                    & $1.684 $              \\
$\alpha_{\rm{over}}$                     & \textunderscore                                  & $0.146 $             \\ \hline
\end{tabular}
}
\end{table}

The non-seismic stellar properties obtained with a Levenberg-Marquardt minimisation technique lay within the $1-\sigma$ errors of all the models and spectroscopic data obtained in \cite{Lillo-Box2014}. By fitting only the lowest frequency modes, we ensured a limited impact of the surface effect correction on the final results, while avoiding additional parameters for the modelling. The best model found lays in a valley of minima that was also found by \cite{Lillo-Box2014}, with a direct fitting of the individual frequencies proving the robustness and consistency of the two asteroseismic modelling approaches. Overall, this model shows a good agreement with both the non-seismic stellar properties and the asteroseismic properties.
\\~\\We note that the Echelle diagram shown in Fig.~\ref{fig._echelle_diagram} includes the empirical surface corrections of \cite{Ball2014}   calibrated from acoustic oscillations of main-sequence stars and added posteriori to the stellar modelling on the $\ell=0$ and the p-dominated $\ell=1$ modes. The surface corrections were only used for display purposes once our final model was obtained, and they were not included in our detailed modelling to avoid any mis-modelling attributed to surface effects.

\section{Kepler-91b}\label{sec_KEP91b_presentation}

In this section, we briefly present the properties of our targeted planet, namely Kepler-91b, which is the only planet known to date orbiting Kepler-91 \cite{Lillo-Box2014,Lillo-Box2014b}. Three authors successfully characterised the planet Kepler 91b, namely \cite{Lillo-Box2014,Lillo-Box2014b,   Sato2015} and \cite{Barclay2015}.We now detail and explain the choice of the planetary parameters of Kepler 91b adopted in this study. While \cite{Lillo-Box2014,Lillo-Box2014b} and \cite{Barclay2015} found very similar system parameters, \cite{Sato2015} found a different system with a lighter planet and a much smaller $a/R_{\star}$. Each author used different sets of radial velocities obtained from their respective instruments and the \textit{Kepler} light curves. Assuming that the orbital period found by the different authors of $6.24668005\pm 0.00002647$ days is valid, one can derive, using Kepler’s third law, the radius of the host star needed to fulfil both constraints on the semi major axis: the period and the measured $a/R_{\star}$. With the stellar mass derived from our modelling, we found that the stellar radius necessary to fulfill both constraints on the semi-major axis is $R_\star= 6.92\pm0.14 R_\odot$ for the parameters of \cite{Sato2015} and $R_\star= 6.37^{+0.88}_{-0.37} R_\odot$ for the parameters derived by \cite{Lillo-Box2014}. Since our stellar model has a radius of $R_\star=6.299 R_\odot,$ we choose to not use the planetary parameters of \cite{Sato2015} due to the impossibility to fit both the period and $a/R_{\star}$. Our choice was also guided by the stellar modelling and characterisation, which is absent from the study of \cite{Sato2015}.
\\~\\We thus chose to adopt the planetary and orbital parameters obtained by \cite{Lillo-Box2014} in our analysis since the uncertainties encompass the parameters obtained in \cite{Barclay2015} and the analysis in \cite{Lillo-Box2014b} is only based on radial velocity measurements.
The adopted characteristics of the system for the rest of the study are presented in Table~\ref{table_planetary_properties}.
\begin{table}[h]
\centering
\caption{Planetary properties taken from \cite{Lillo-Box2014} and adopted in this article. }\label{table_planetary_properties}
\begin{tabular}{lr}
\hline\hline
\multirow{2}{*}{Parameters} & \multirow{2}{*}{Adopted values}  \\
                                    &                                                                                 \\ \hline
$M_{\rm{pl}}$ {[}$M_{\rm{Jup}}${]}               & $0.88^{+0.17}_{-0.33}$                                         \\
$R_{\rm{pl}}$ {[}$R_{\rm{Jup}}${]}              & $1.384^{+0.011}_{-0.054}$                                          \\
$a/R_{\star}$        & $ 2.45^{+0.15}_{-0.30}$                                     \\
Period [days]           & $6.24668005\pm 0.00002647$                                       \\
$e$    & $0.066^{+0.0013}_{-0.0017}$                                               \\
$i$ {[}deg{]}                     & $68.5^{+1.0}_{-1.6}$                                          \\
\hline
\end{tabular}
\end{table}
\\~\\Despite its unusually small inclination and high-impact parameter (\cite{Sato2015, Barclay2015}), which would normally have resulted in a non-transiting planet, Kepler-91b is still observable thanks to the inflated radius of its host star.
Regarding stellar rotation, \cite{Lillo-Box2014} derived a surface rotation using spectroscopy and a global spectrum fit based on a grid of parameters, the surface rotation obtained is $v\sin i=6.8 \pm 0.2\ \rm{km s}^{-1}$ or $\Omega_\star=265^{+18}_{-16}$ nHz (corresponding to a period of $44^{+2}_{-3}$ days) using the inclination, stellar radius, and uncertainties given in \cite{Lillo-Box2014}. Excluding this rotation from being the result of a previous engulfment, we could reproduce the observed stellar surface rotation using our orbital evolution model with an initial surface rotation at the beginning of the MS of $\Omega_\star=0.4\ \Omega_{\rm{crit}}$, where $\Omega_{\rm{crit}}$ is the critical rotation rate after which the stellar envelope is not bounded to the star. Even if this rotation can be reproduced, we noticed that the errors on the $v\sin i$  are significantly small compared to similar stars such as Kepler-56 \citep{Huber2013}, and that Kepler-91 is indeed a fast rotator. The surface rotation could also be confirmed with asteroseismology using rotational splittings; however, none of them were clearly identified by \cite{Lillo-Box2014}.

\section{The orbital evolution code}\label{section_code}

In this section, we briefly describe the structure and the physics included in the orbital evolution code used in this work and present the modifications implemented. This orbital evolution code was first introduced in \cite{Privitera2016I, Privitera2016II, Meynet2017,Rao2018} and recently used in \cite{Pezzotti2021}. The principle of this code is to couple the evolution of the planet with the one of the host star by giving the stellar model as input. In particular, stellar models provide quantities such as the stellar momentum of inertia necessary to follow the evolution of the stellar rotation and the dynamical interactions between the star and the planets.
\\~\\In our code, we account for the evolution of the planetary orbit due to the dissipation of tides. The tidal force is resultant of the gravitational force appearing in bodies not described as point-mass objects. The tidal force drives a mass redistribution into bodies by perturbing their gravitational potentials. Tides can be divided into two components: the equilibrium tide, which is a large-scale circulation resulting from hydrostatic adjustment; and the dynamical tides \citep{Zahn1977}, which come from the excitation of eigenmodes of oscillations. A part of the energy used to displace the fluid elements inside the body’s gravitational potential is directly extracted from the orbital energy of the system and the tidally perturbed body's rotation. This energy is then dissipated inside bodies through friction or heat diffusion in radiative zones. The tidal dissipation affects the orbital evolution of a system through the evolution of a few orbital parameters, the semi-major axis, the eccentricity, the obliquity, the precession, and the rotation periods of the involved celestial bodies.
\\~\\In the orbital evolution code, we can study the orbital evolution of single planet around its host star. We assume that the planet is on a coplanar orbit with respect to stellar equatorial plane, and we do not account for the planetary obliquity. In this new version of the code, we extend the study of the evolution of the orbit to planets on eccentric orbits. The code computes the evolution of the planetary semi-major axis caused by the tidal force; gravitational, frictional and drag forces; and migration linked to planetary mass loss \citep{Zahn1966, Zahn1977, Zahn1978, Zahn1989, Villaver2009, Villaver2014}. Initially, our orbital evolution code took into account the equilibrium and dynamical tides dissipated in the stellar convective envelopes, while the dissipation in the planet was not accounted for. Finally, the planetary mass loss is computed using the energy limited formalism \citep{Watson1981, Lammer2003, Pezzotti2021}. We explored the effect's mass loss and found it to be negligible for this target.

\subsection{Equilibrium tides}

A precise description of the dynamical evolution of Kepler 91-b is key to constraining the tidal dissipation coefficients of the planetary equilibrium tides. To that end, and moreover to study the case of eccentric orbits, we decided to implement the evolution of the eccentricity inside our orbital evolution code \citep{Privitera2016I, Privitera2016II, Meynet2017, Rao2018}. Since this planet is on an eccentric orbit, as a good fraction of the hot Jupiters observed around RGB stars are, we decided to implement the evolution of the orbital eccentricity in our orbital evolution code.
\\~\\From the formalism of the stellar equilibrium tides included in our orbital evolution code (see \cite{Villaver2009, Villaver2014}), derived from that of \cite{Zahn1966, Zahn1977, Zahn1978, Zahn1989}, we generalised the expression of the semi-major-axis variation to include the contribution of the eccentricity. In the same formalism, the variation of eccentricity was obtained directly from the expressions of \cite{Zahn1966, Zahn1977, Zahn1978, Zahn1989}. The expressions of the semi-major axis and eccentricity evolution only accounting for the equilibrium tides raised in the star are as follows:

\begin{multline}\label{eq._a_dot_star}
\left(\frac{1}{\rm{a}}\frac{\rm{d} a}{\rm{d} t} \right)_{\rm{eq.tides},\star}  = \\ \frac{f_{\rm{orb}}}{\tau}\frac{M_{\rm{env}}}{M_{\star}}q(1+q)\left(\frac{R_\star}{\rm{a}}\right)^8 \left[ N_{a2}(\rm{e})\frac{\Omega_\star}{n} - N_{a1}(\rm{e}) \right],
\end{multline}
\begin{multline}\label{eq._e_dot_star}
\left(\frac{1}{\rm{e}}\frac{\rm{d} e}{\rm{d} t} \right)_{\rm{eq.tides},\star} = \\ \frac{11}{4} \frac{f_{\rm{orb}}}{\tau} \frac{M_{\rm{env}}}{M_{\star}}q(1+q)\left(\frac{R_\star}{\rm{a}}\right)^8 \left[ \Omega_{e}(\rm{e})\frac{\Omega_\star}{n} - \frac{18}{11} N_{e}(\rm{e}) \right],
\end{multline}
\\~\\with 
\begin{equation}
N_{a2}(e)=\frac{1+\frac{15}{2}e^2 +\frac{45}{8}e^4 +\frac{5}{16}e^6}{(1-e^2)^{6}},
\end{equation}
\begin{equation}
N_{a1}(e)=\frac{1+\frac{31}{2}e^2 +\frac{255}{8}e^4 +\frac{185}{16}e^6 +\frac{25}{64}e^8}{(1-e^2)^{15/2}},
\end{equation}
\begin{equation}
\Omega_{e}(e)=\frac{1+\frac{3}{2}e^2 +\frac{1}{8}e^4}{(1-e^2)^{5}},
\end{equation}
\begin{equation}
N_{e}(e)=\frac{1+\frac{15}{4}e^2 +\frac{15}{8}e^4 +\frac{5}{64}e^6}{(1-e^2)^{13/2}},
\end{equation}
\\~\\where $\rm{a}$ denotes the semi major axis of the planet, $e$ its eccentricity, $M_{\rm{env}}$ its mass, $M_{\star}$ the mass of the star, $M_{env}$ the mass of its envelope, $R_\star$ its total radius, $q$ is the planet-to-star mass ratio ($q=\frac{M_{\rm{pl}}}{M_{\star}}$), $\Omega_\star$ the stellar surface rotation rate, and $n$ the planet orbital frequency.$\tau$ corresponds to the eddy turnover timescale \citep{Rasio1996, Villaver2009} defined as
\begin{equation}
\tau=\left[ \frac{M_{\rm{env}}(R_{\star}-R_{\rm{env}})^2}{3L} \right]^{\frac{1}{3}},
\end{equation}
\\~\\with $R_{\rm{env}}$ being the radius at the base of the convective envelope, and L the bolometric luminosity.
\\Finally, $f_{\rm{orb}}$ is a numerical factor equal to $1$ when $\tau< \frac{P_{\rm{orb}}}{2}$ (with $P_{\rm{orb}}$ being the orbital period of the planet); otherwise, $f_{\rm{orb}}$ is expressed as
\begin{equation}\label{eq._f}
f_{\rm{orb}}=\left( \frac{P_{\rm{orb}}}{2\tau}\right)^2.
\end{equation}
\\~\\The equilibrium tides are only considered when the star has a convective envelope. The term $N_{a2}(e)\frac{\Omega_\star}{n} - N_{a1}(e)$ in Eq.\ref{eq._a_dot_star} defines whether the planet is inside the co-rotation radius, the radius at which the orbital period of the planet is equal to the stellar rotation period. The impact of tides is a widening of the orbit when the planet is beyond the co-rotation radius and shrinkage when the planet is inside it.
\\~\\The equilibrium tides raised inside the planets are also included in the code with the constant time lag formalism \citep{Darwin1879}. Assuming a planet in pseudo-synchronisation \citep{Hut1981} with a rotation rate of $\omega_{\rm{pl}}=\omega_{\rm{pl, ps}}$, the expressions for the evolution of the semi major axis and eccentricity caused by planetary equilibrium tides are 
\begin{multline}\label{eq._a_dot_pl}
\left(\frac{1}{\rm{a}}\frac{\rm{d} a}{\rm{d} t} \right)_{\rm{eq.tides},\ \rm{pl} }=\\  6 k_{\rm{2, p}}\Delta t_{\rm{p}} \frac{1}{ q} \left( \frac{R_{\rm{pl}} }{\rm{a}} \right)^5 n^2 \left[ N_{a2}(\rm{e})\frac{\omega_{\rm{pl, ps}}}{n}- N_{a1}(\rm{e}) \right],
\end{multline}
\begin{multline}\label{eq._e_dot_pl}
\left(\frac{1}{\rm{e}}\frac{\rm{d} e}{\rm{d} t} \right)_{\rm{eq.tides},\ \rm{pl} } = \\ \frac{33}{2} k_{\rm{2, p}}\Delta t_{\rm{p}} \frac{1}{q}\left( \frac{R_{\rm{pl}} }{\rm{a}} \right)^5 n^2 \left[ \Omega_{e}(\rm{e})\frac{\omega_{\rm{pl, ps}}}{n} - \frac{18}{11} N_{e}(\rm{e}) \right],
\end{multline}
where $R_{\rm{pl}}$ denotes the radius of the planet, $k_{\rm{2, p}}$ is the potential Love number of degree 2 of the planet, and  $\Delta t_{\rm{p}}$ its constant time lag. 
For a rotationally synchronised planet on a circular orbit, the tidal dissipation inside the planet vanishes even if a whole spectrum of frequencies, not considered in our code, could dissipate energy. 
\\Following \cite{Hut1981}, in the case of a planet in a state of tidal locking, the ratio $\frac{\omega_{\rm{pl, ps}}}{n} $ can be expressed as
\begin{equation}
\frac{\omega_{\rm{pl, ps}}}{n}=\frac{1+\frac{15}{2}e^2 + \frac{45}{8}e^4+\frac{5}{16}e^6 }{(1+3e^2+\frac{3}{8}e^4)(1-e^2)^{3/2}}.
\end{equation}
\\~\\The tides raised in the planet can dissipate energy in the considered body through friction. In the case of a pseudo-synchronised planet, the energy dissipated inside the planet $\dot{E}_{\rm{pl}}$ \citep{Hut1981} is given by
\begin{equation}\label{eq._E_dot}
\dot{E}_{\rm{pl}}= 6 k_{\rm{2, p}}\Delta t_{\rm{p}} \frac{GM_\star^2}{R_{\rm{pl}}}\left( \frac{R_{\rm{pl}}}{a} \right)^6n^2\left(N_{a1}(e)-\frac{N_{a2}(e)^2}{\Omega(e)}\right),
\end{equation}
where $\Omega(e)$ is defined as 
\begin{equation}
\Omega(e)=\frac{1+3e^2 +\frac{3}{8}e^4}{(1-e^2)^{9/2}}.
\end{equation}
\\~\\This dissipated energy is then added to the energy balance of the planet, assuming that all the tidal flux is instantly and homogeneously deposited in the planet (see Eq.~\ref{eq._Teq}).
\\~\\Although in our orbital evolution code the dissipation of dynamical tides in stellar convective zones is taken into account when the planet is on a circular and coplanar orbit with respect to the stellar equatorial plane, we do not take them into account here. The formalism adopted in our code \citep{Ogilvie2013} does not apply to the case of dynamical tides raised in the star by planet on eccentric orbits. The dynamical tides were not expected to play a key role for this particular system during the stellar MS as the dissipation of stellar tides is negligible compared to the planetary tides in the regimes considered here.
\\~\\The total change of semi-major axis caused by tides is thus given by 
\begin{multline}
\left(\frac{1}{\rm{a}}\frac{\rm{d} a}{\rm{d} t} \right)_{\rm{tides}} = \left(\frac{1}{\rm{a}}\frac{\rm{d} a}{\rm{d} t} \right)_{\rm{eq.tides},\star} + \left(\frac{1}{\rm{a}}\frac{\rm{d} a}{\rm{d} t} \right)_{\rm{eq.tides},\ \rm{pl} },
\end{multline}
\\~\\and the same expression can be adopted for the change of eccentricity: 
\begin{multline}
\left(\frac{1}{\rm{e}}\frac{\rm{d} e}{\rm{d} t} \right)_{\rm{tides}}= \left(\frac{1}{\rm{e}}\frac{\rm{d} e}{\rm{d} t} \right)_{\rm{eq.tides},\star} + \left(\frac{1}{\rm{e}}\frac{\rm{d} e}{\rm{d} t} \right)_{\rm{eq.tides},\ \rm{pl} }.
\end{multline}

\subsection{Computation of the stellar rotation}
As specified in the introduction, the orbital evolution code has the responsibility of coupling the stellar evolution with the planet evolution, and in particular computing the stellar rotation. In this study, we assumed that the star rotates as solid body along its evolution; this assumption is supported by helioseismic measurements as well as by asteroseismic observations of the internal rotation of solar-type stars.
\\~\\The initial value of the stellar surface rotation rate is given as input in the orbital evolution code and is chosen in order to reproduce the surface rotation rate observed for Kepler-91. The braking of the stellar surface by magnetized winds is neglected for this target, since it is expected to be inefficient for a star with a mass similar to the one of Kepler-91, which exhibits a very shallow convective envelope \cite{Matt2012}. Using the conservation of the angular momentum of the tidal force, the evolution of the specific angular momentum of the star $J_\star$ resulting from the migration and circularisation of the planet can be expressed as 
\begin{multline}
\frac{\rm{d} J_\star}{\rm{d} t}=\\ -\left(G\mu a (1-e^2)\right)^{\frac{1}{2}}\left[\frac{1}{2}\left(\frac{1}{a}\frac{\rm{d} a}{\rm{d} t} \right)_{\rm{tides}} - \frac{e^2}{1-e^2} \left(\frac{1}{e}\frac{\rm{d} e}{\rm{d} t} \right)_{\rm{tides}}\right], 
\end{multline}
where $\mu$ denotes the reduced mass of the system, expressed as 
\begin{equation}
\frac{1}{\mu}= \frac{1}{M_{\star}} + \frac{1}{M_{\rm{pl}}}.
\end{equation}

\subsection{Evolution of the planetary radius}

\begin{figure}[h]
\centering
\includegraphics[width=\hsize]{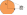}
\caption{Diagram underlying the geometric increase of irradiation of the host star onto the planet surface caused by the increase of the stellar radius. }
\label{fig._irradiation_schema}
\end{figure}

The particular case of close-in planets orbiting RGB stars requires several adjustments on the modelling of the planet and the treatment of the planetary mass loss.
The planet equilibrium temperature should be adjusted to account for the increase of the irradiation received from the host star over the planetary surface, caused by the increase of the stellar radius. This situation is illustrated in  Fig.~\ref{fig._irradiation_schema}. Physically, because of the increase of the stellar radius during the RGB phase, a larger portion of the planetary surface is irradiated by the star, and consequently its equilibrium temperature increases.
For the case of Kepler-91, \cite{Lillo-Box2014} found that the total surface of Kepler 91b irradiated by the star is above $70\%,$ showing that this mechanism is non-negligible for close-in planets around RGB stars. In Fig.~\ref{fig._irradiation_schema}, the key parameter $\alpha$ denotes the angle between a situation with a normal planetary irradiation and the situation taking into account the geometrical increase of irradiation of the host star onto the planet surface. Using geometry, the expression of  $\alpha$ is given by 

\begin{multline}
\alpha=\sin^{-1}\left( \frac{R_\star}{\sqrt{a^2+R_\star^2}} \right) - \sin^{-1}\left( \frac{R_{\rm{pl}}}{\sqrt{a^2+R_\star^2}} \right) \\ \approx \sin^{-1}\left( \frac{R_\star}{\sqrt{a^2+R_\star^2}} \right).
\end{multline}

This angle is non-negligible for close-in planets around giant stars, or at least when the stellar radius has the same order of magnitude as the orbital separation. For planets with wide orbits, $\alpha\sim 0$ and the planet recovers its $50\%$ of irradiated surface. To compute the irradiation of the planet, we introduced a quantity $\beta$ defined as the ratio of the actually surface irradiated to the surface irradiated without these geometrical considerations. This coefficient can be expressed as
\begin{equation}
\beta= \frac{1}{2\pi R_{\rm{pl}}^2}\int^{2\pi}_0 \rm{d}\phi \int^{\frac{\pi}{2}+\alpha}_{0}  R_{\rm{pl}}^2 \sin(\theta)\rm{d}\theta  =1+ \sin (\alpha).
\end{equation}
\\~\\Finally, the equilibrium temperature $T_{\rm{eq.}}$ of the planet including the geometrical correction and including the energy dissipated through planetary tides, presented in Equation~\ref{eq._E_dot}, can be expressed as
\begin{equation}\label{eq._Teq}
T_{\rm{eq.}}= \left(\frac{L_{\star} (1-A) \beta }{16\pi a^2 \sqrt{1-e^2} \sigma_{SB} }+ \frac{\dot{E}_{\rm{pl}}}{4\pi R_{\rm{pl}}^2\sigma_{SB}}\right)^{\frac{1}{4}},
\end{equation}
\\~\\where $L_{\star}$ denotes the stellar luminosity, $\sigma_{SB}$ the Stefan–Boltzmann constant, and $A$ the planetary bond albedo. In the previous formula, $\beta$ is treated as a simple coefficient equal to $1$ for planets with wide orbits; thus, this expression is a generalisation for the classical equilibrium temperature computation. For each of the evolutionary tracks presented below, the bond albedo was kept at 0.30, the maximum observed albedo comes from the sample of hot Jupiters of \cite{Adams2022}.
\\~\\Determining the mechanisms that concur in the inflation of the planetary radius for hot Jupiters represents a challenging task. The high quantity of energy deposited in the atmosphere leads to a significant increase in the planetary radius impacting the dynamical evolution and the planetary mass loss. In our cases, the classic mass-radius relations can be inaccurate and underestimate the planetary radius. Moreover, the radii of hot and ultra-hot Jupiters are dominated by their irradiations more than their masses; thus, we chose to compute the planetary radius with the temperature-radius law presented in \cite{Owen2018} only valid for planets with masses greater than $0.2M_{\rm{Jup}}$. The expression of the initial radius of the planet given in \cite{Owen2018} is
\begin{equation}\label{eq._planet_radius}
R_{\rm{pl}} =0.8 f_{\rm{r}} \left( \frac{T_{\rm{eq.}}}{ 1100 K}\right)^{0.7224} R_{\rm{Jup}} ,
\end{equation}
with $f_{\rm{r}}$ being a free parameter with a value laying between $1$ and $1.5$; the parameter is adjusted in our code to reproduce the observed planetary radius. Following \cite{Owen2018}, the planetary radius evolves according to
\begin{equation}\label{eq._planet_radius_dot}
\dot{R}_{\rm{pl}} =0.57792f_{\rm{r}}\frac{\dot{T}_{\rm{eq.}}}{(1100K)^{0.7224}T_{\rm{eq.}}^{0.2776}}R_{\rm{Jup}}  , 
\end{equation}
where $\dot{T}_{\rm{eq.}}$ denotes the time derivative of the planetary equilibrium temperature.

\section{Orbital evolution of Kepler-91b}\label{sec_KEP91b_modelling}

In this Section, different observational constraints of the Kepler-91 system are used to constrain the physics of tides and their dissipation. The study of the Kepler-91 system is divided into three distinct parts. First, in Section~\ref{subsec_dissipation_star_planet}, we start by carrying out a theoretical comparison between the the dissipation of the planet and of the star. Then, we constrain the physics of tides and their dissipation using different observational constraints on the system. Section~\ref{subsec_evolution_dissipation} is dedicated to the study of the past evolution of Kepler-91b to quantify and constrain the dissipations of tides in the planet during this period of time. In Section~\ref{subsec_evolution_dissipation_star}, we use the constraints obtained on the planetary equilibrium tides to quantify the dissipation of the stellar equilibrium tides through the different phases of stellar evolution, both for the past and future evolution of the system.

\subsection{Comparison of the dissipation in the planet and that in the star}\label{subsec_dissipation_star_planet}
\begin{figure}[h]
\centering
\includegraphics[width=\hsize]{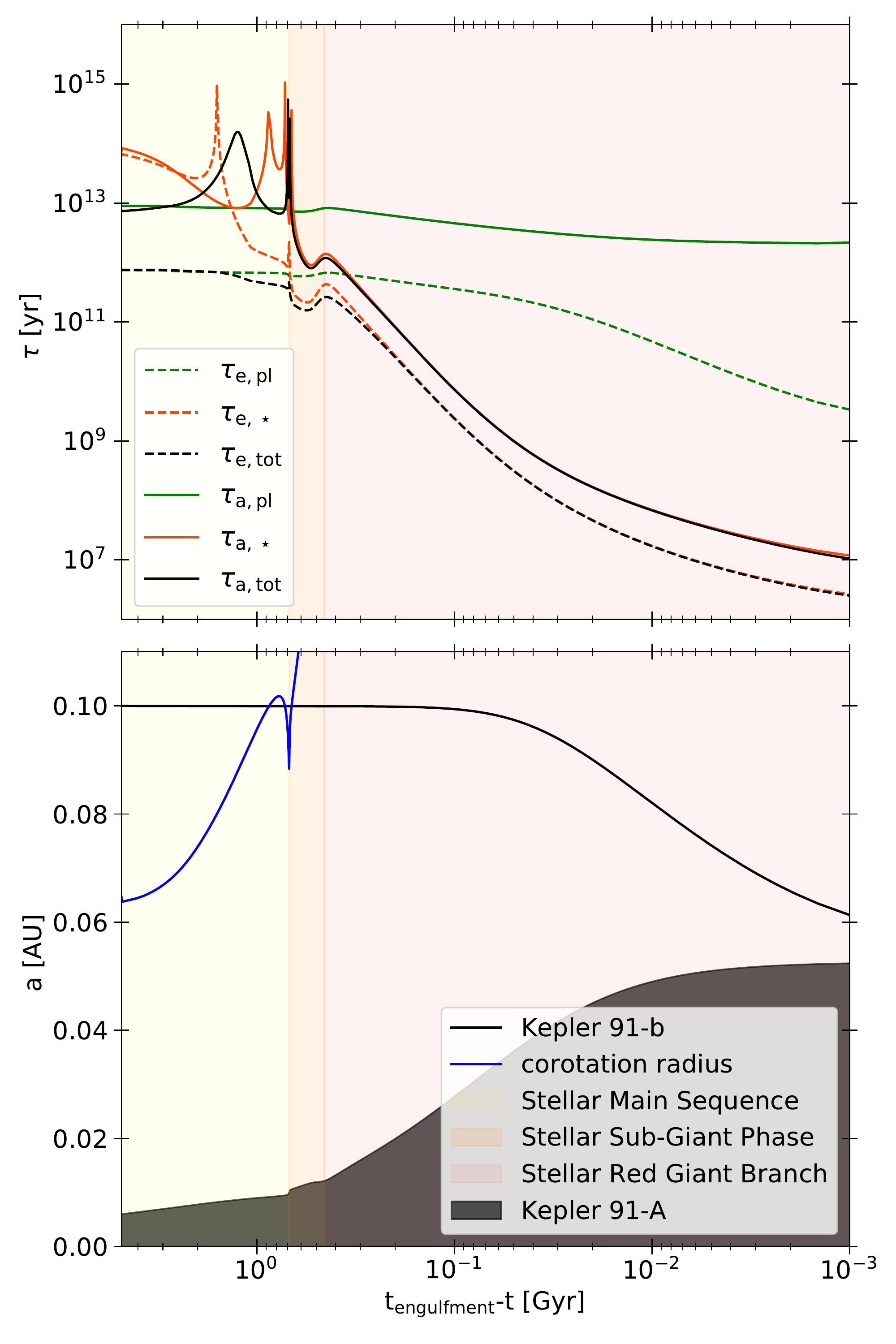}
\caption{Illustration of the impact of the equilibrium planetary and stellar tides on the orbital evolution of a system represented through the timescales of migration $\tau_a=(a/\dot{a})$ and circularisation $\tau_e=(e/\dot{e})$. The system evolves around the stellar model of Kepler 91 presented in Sect.~\ref{sec_KEP91b_modelling}. The initial mass of the planet is set at $0.88$ Jupiter's mass, its initial semi major axis is set at $0.10\ \rm{AU}$, its initial eccentricity to $0.2$ and $k_{\rm{2, p}}\Delta t_{\rm{p}} $ is set at $10^{-2}$ s. The upper panel illustrates the different timescales, the dashed lines correspond to the circularisation timescales while the solid lines represent the migration timescales. Each modified component of these timescales (planetary, stellar and their total) are represented namely in green, red and black. The lower panel correspond to the associated migration of the planet, the engulfment of the planet occurs at $t_{\rm{engulfment}}=4.897$ Gyr. Finally the shaded areas correspond to the different evolutionary phases of the star, in yellow the MS, in orange, the sub-Giant phase, and in red the RGB phase.}
\label{fig._timescale_eq.tides}
\end{figure}

To illustrate the variation of the different components of the tidal dissipation, we carried out an initial comparison of the migration timescale $\tau_a$ and circularisation timescale $\tau_e$. Figure~\ref{fig._timescale_eq.tides} illustrates the evolution of these timescales over the lifetime of the whole system, only accounting for equilibrium tides. 
\\~\\Figure~\ref{fig._timescale_eq.tides} shows that during the MS of the star (ending at an age of $\sim 4.5$ Gyr), the circularisation is totally dominated by the planetary tides in this case. The slow migration of the planet is led both by the stellar and planetary tides during the MS, while during the stellar sub-giant phase and RGB phase the stellar tides dominate. The particular peaks seen in $\tau_{\rm{a}, star}$, $\tau_{\rm{tot}, star}$ and $\tau_{\rm{e}, star}$ during the MS are caused by the co-rotation radius almost being reached. During the stellar RGB phase, the dissipation and the engulfment of the planet are driven by the stellar tides. We also found that the circularisation acts on shorter timescales than the migration during the stellar MS, while during the engulfment both timescales have the same order of magnitude even if the circularisation still happens faster.
Concerning the stellar tides, all the results presented were expected \citep[see][]{Villaver2014, VanEylen2016, Grunblatt2018}, confirming the accuracy of our orbital evolution code. 

\subsection{Constraints on the maximum value of the planetary tidal dissipation factors}\label{subsec_evolution_dissipation}

\begin{figure}[h]
\centering
\includegraphics[width=\hsize]{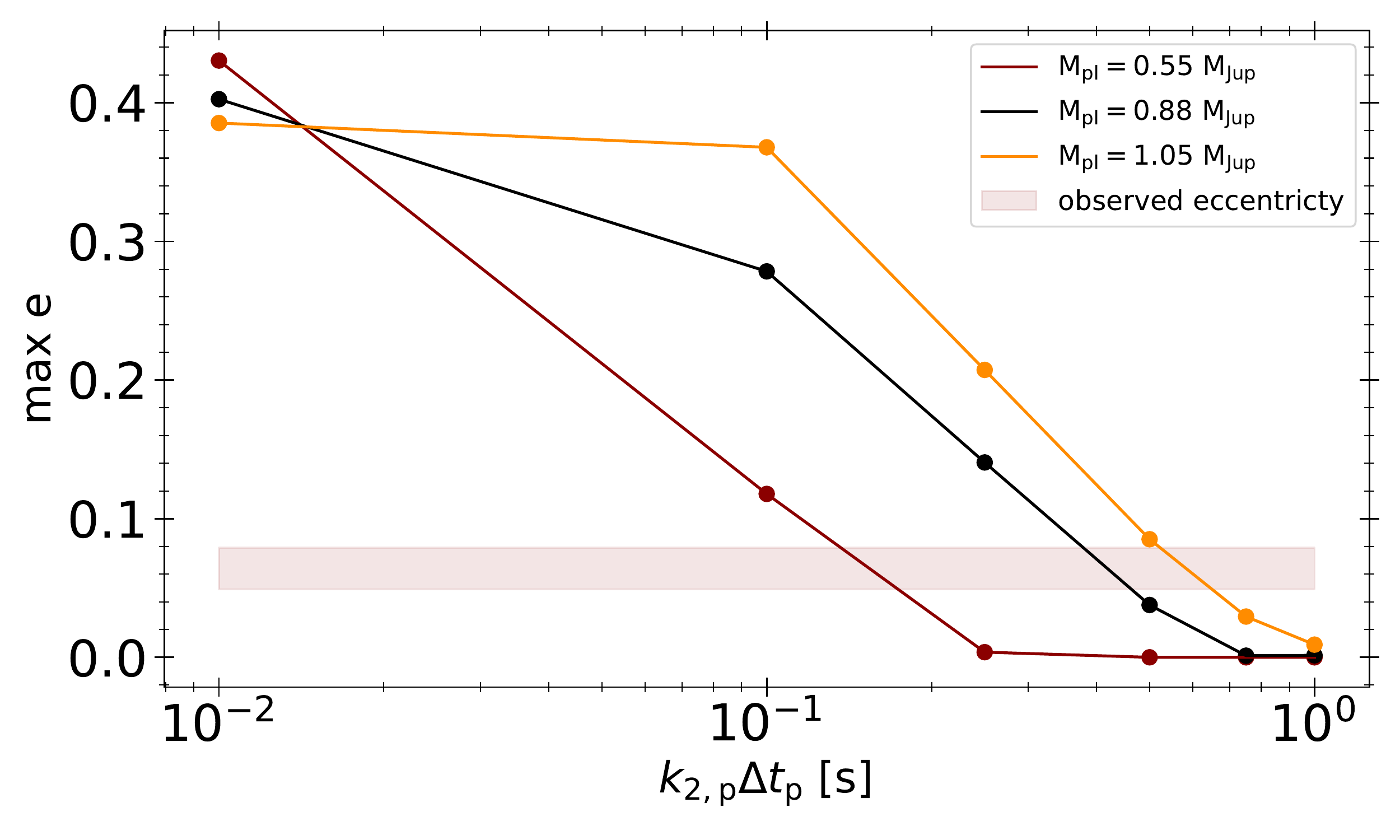}
\caption{Results from the grid of evolutionary tracks illustrating the maximum eccentricity that Kepler 91b can reach for given values of the tidal dissipation parameter $k_{\rm{2, p}}\Delta t_{\rm{p}}$. These results are compared to the observed eccentricity for different planetary masses considered.  A summary of the initial and final stellar,  orbital and planetary parameters of each evolutionary track used here is presented in Tab.~\ref{table_dissipation_constraint}. }
\label{fig._dissipation_k2dt}
\end{figure}

\begin{figure}[h]
\centering
\includegraphics[width=\hsize]{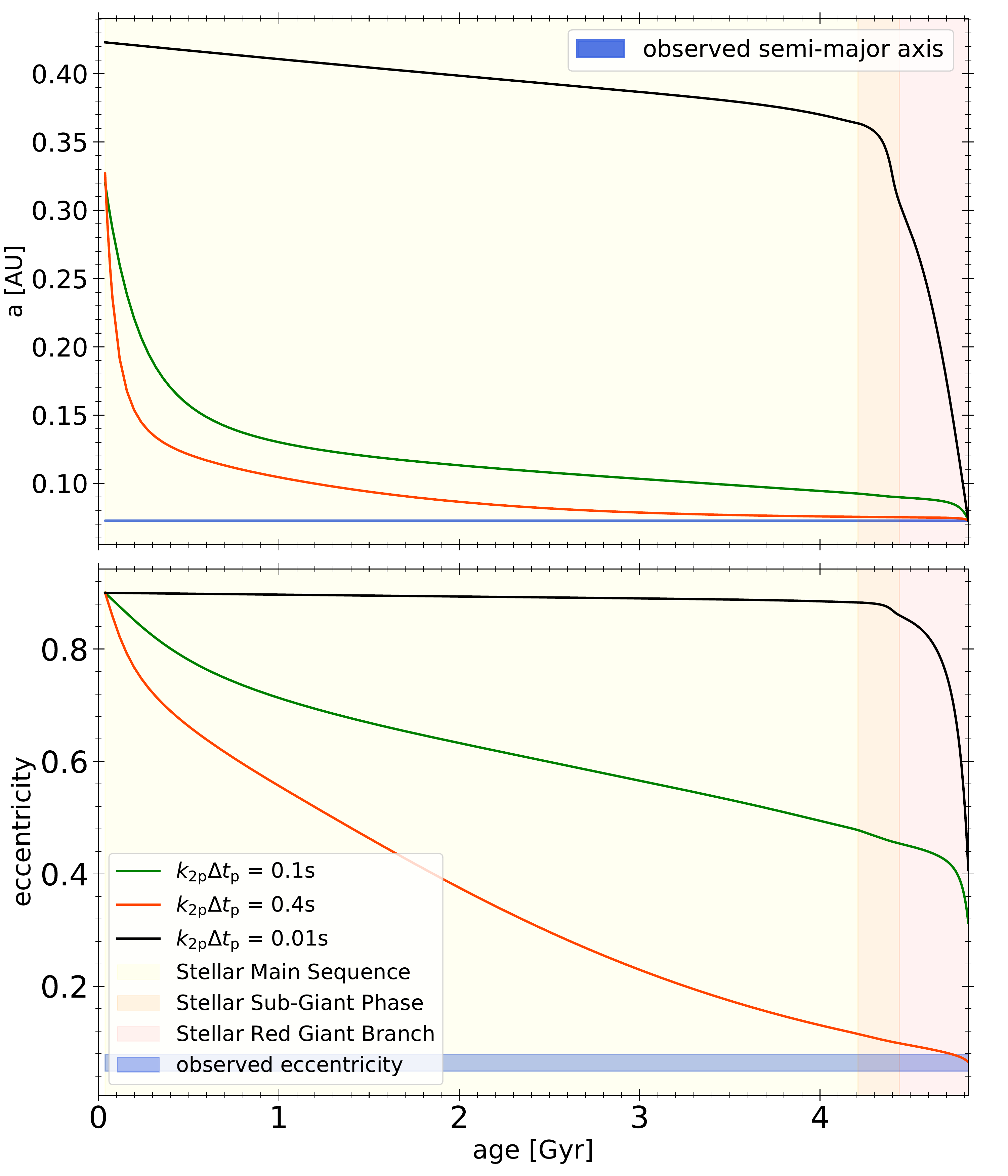}
\caption{Evolution of the semi major axis and eccentricity as a function of time on, respectively, the top and bottom panel. Each curve represents the path taken by the planet with the different $k_{\rm{2, p}}\Delta t_{\rm{p}}$ used in Fig.~\ref{fig._dissipation_k2dt}. The blue shaded areas are to the corresponding observed quantities. The yellow area corresponds to the stellar MS, the orange one is the stellar sub-Giant phase, and the red one is the stellar RGB phase. The age of the system corresponds to the end of these evolutionary sequences. }
\label{fig._evolution}
\end{figure}

In this section, we study the past evolution of Kepler-91b to constrain the physics of tides and their dissipation. Before presenting our method, we should remind the reader of the different assumptions made during this study. First, on the stellar side, all the uncertainties presented here do not include possible uncertainties on the stellar modelling done in Sect.~\ref{sec_KEP91b_presentation}. Moreover, we ensure that the observed stellar rotation of $\Omega_\star=265^{+18}_{-16}$ nHz is reproduced in our modelling at the age of the system. On the planetary side, we adopted the planetary characteristics presented in Tab.~\ref{table_planetary_properties}; each evolutionary track starts at the beginning of the stellar MS to avoid any uncertainties coming from the dynamical tides. The planet bond albedo was kept constant during the whole evolution with $A=0.30,$ and we also assumed that the planet was in pseudo-synchronisation throughout its evolution, meaning that the dissipation of the planetary equilibrium tides can be underestimated at the beginning of the evolution. Finally, we assumed that the planet's orbital plane was co-planar with respect to the stellar equatorial plane and that the system is only composed of one planet during its whole lifetime.
\\~\\To constrain the dissipation of the planet equilibrium tides, we built a grid of evolutionary tracks with different planetary equilibrium tides' dissipation coefficients: $k_{\rm{2, p}}\Delta t_{\rm{p}}$.  The properties and initial parameters used to compute the grid are resumed in Tab.~\ref{table_dissipation_constraint}.  The end of each evolutionary track corresponds to the age of the system given in Tab.~\ref{table_stellar_properties}. The purpose of this grid is to find the maximum eccentricity that Kepler-91b can reach at the age of the system given a tidal dissipation factor while fitting all the other planetary and stellar properties. The initial semi major axis is optimised in order to reproduce the observed period at the age of the system and for each $k_{\rm{2, p}}\Delta t_{\rm{p}}$.  
The maximum eccentricity reached by the system allows us to constrain the tidal dissipation of the planet by giving a limit to $k_{\rm{2, p}}\Delta t_{\rm{p,}}$ after which point the observed eccentricity cannot be reproduced.
\\~\\It is  possible to notice that a degeneracy exists in between the initial parameters (orbital, stellar) and the value of the tidal dissipation factor in reproducing the current properties of the system. However, this degeneracy does not concern the tidal dissipation in the host star, which is computed by following the structural and rotational evolution of the host star,  as in Eqs. ~\ref{eq._a_dot_star} and \ref{eq._e_dot_star}.  Therefore, while  it is possible to tune the value of the dissipation factor for the planet, this is not the case for the tides dissipated in the host star.  In that sense, the degeneracy on the stellar tidal dissipation factor is removed. 
There is a degeneracy in the initial orbital and planetary parameters and the planetary tidal dissipation factor, but not on the tidal dissipation factor of the host star.  For this reason,  we performed a study of the parameter space and found some constraints on $k_{\rm{2, p}}\Delta t_{\rm{p}}$.
Instead of providing a set of degenerate initial parameters allowing us to reproduce the observations, we looked for the maximum value of the tidal dissipation factor that,  by fine tuning of the rest of the parameters,  can lead to the reproduction of the system properties.
\\~\\ We should mention that we are not trying to reproduce the observed eccentricity here.  We are looking at the maximum eccentricity that the planet can have at the end of an evolutionary track while reproducing all the others orbital parameters (i.e. the period, planet radius, and stellar rotation). 
The initial eccentricities were set at the maximum values tolerated by the system to avoid any early engulfment, or to 0.90 to avoid any non-physical behaviour close to the singularity of e=1.0. With this fixed initial eccentricity and each dissipation factor, we experimentally adjusted the initial semi-major axis to reproduce the planet period at the current age of the system.  We define a limit track for the planet orbital parameters with a given dissipation factor. 
\\~\\At the age of the system,  two possible scenarios can occur with respect to the final eccentricity value of the planet orbit: 
 \\If the eccentricity is higher than the observed one within the uncertainties, it is possible to adjust the initial semi-major axis and eccentricity to reproduce the observed planetary parameters.
\\ If the eccentricity is lower than the observed one, it is not possible to find a combination of initial parameters to reproduce the observed orbital and planetary parameters; thus, the dissipation factor is not consistent with the observations.
\\Particular care was taken to reproduce the observed planetary radius to not bias the tidal dissipation. Finally, we made this grid for three different planetary masses, M$_{\rm{pl}}=0.88^{+0.17}_{-0.33}\ $M$_{\rm{Jup}}$, to account for the uncertainties on the mass of the planet. The results of this model grid and the consequent constraints on the tidal dissipation coefficient are presented in Fig.~\ref{fig._dissipation_k2dt} and Tab.~\ref{table_dissipation_constraint}.
\\~\\Figure~\ref{fig._dissipation_k2dt} shows that when taking into account both the uncertainties on the mass of the planet and on its observed eccentricity, the maximum value of $k_{\rm{2, p}}\Delta t_{\rm{p}}$ allowed to reproduce the eccentricity is $\sim 0.4\pm0.25$ s. For such values, the orbital migration undergone by the planet is quite significant, starting at an orbital distance of $0.33$ AU with an eccentricity of 0.90;  the planet finishes its evolution at $0.0721$ AU with an eccentricity of $0.049$. The damping of the eccentricity mostly results from the planet equilibrium tides during the early time of the system in this scenario.
\\~\\Figure~\ref{fig._dissipation_k2dt} also illustrates two regimes of past eccentricity damping, with the first one led by the planet equilibrium tides, where the past damping is proportional to $1/q$ ($q=M_{\rm{pl}}/M_{\star}$), as on the right side of Fig.~\ref{fig._dissipation_k2dt}. The second regime corresponds to a domination of the past damping by the stellar equilibrium tides. This regime occurs in the Kepler-91 system when $k_{\rm{2, p}}\Delta t_{\rm{p}}\lesssim 0.015$ s. This transition in the regimes of dissipation can be seen with the transition in the dependency of $q$ in Fig.~\ref{fig._dissipation_k2dt}, as stellar equilibrium tidal dissipation is proportional to $q^2$ (see Eqs.~\ref{eq._a_dot_star} and \ref{eq._e_dot_star}), while the planet equilibrium tides dissipation is proportional to $1/q$ (see Eqs.~\ref{eq._a_dot_pl} and \ref{eq._e_dot_pl}).
\cite{Leconte2010} showed that for such systems, the planet tidal dissipation factor is of the order of $k_{\rm{2, p}}\Delta t_{\rm{p}} \sim 10^{-2}-10^{-3}$ s, meaning that in our case we expect the stellar equilibrium tides to have dominated the past evolution of Kepler 91b while planet equilibrium tides would have had a limited impact.
\\~\\~\\To express the dissipation of the planetary tides, we chose to use the constant time lag (CTL) model.  However, one may choose to adopt the constant phase lag (CPL) model with the $Q'$ formalism \citep{Goldreich1963, Goldreich1966}. Staring from our Eqs.~\ref{eq._a_dot_pl} and \ref{eq._e_dot_pl}, we can estimate the dissipation in the CPL model by assuming that the planet is in pseudo-synchronisation, as we did. In this case, the most dissipative tides are the annual eccentric tides, and we can link the dissipation of tides in the CPL model to their dissipation in the CTL model using \citep{Leconte2010}:
\begin{equation}\label{eq._Q_prime}
Q'^{-1}_{\rm{p}}\approx\frac{2}{3}k_{\rm{2, p}}\Delta t_{\rm{p}} n,
\end{equation}
where $Q'_{\rm{p}}$ is the reduced tidal quality factor of the planet quantifying the efficiency of the tidal dissipation. $Q'_{\rm{p}}$ is calibrated depending on the type of planets studied and kept constant for the whole evolution. We should mention that the CPL model is expected to be accurate to $e^2$ \citep{Leconte2010}; therefore, one should be careful before using it for very eccentric orbits.
\\~\\Equation~\ref{eq._Q_prime} can be used to replace $k_{\rm{2, p}}\Delta t_{\rm{p}}$ in Eqs.~\ref{eq._a_dot_pl} and \ref{eq._e_dot_pl}, with a new free parameter controlling the dissipation of tides: $Q'_{\rm{p}}$. Repeating the method previously presented on a new grid of model with $Q'_{\rm{p}}$ , we obtain the equivalent of Fig.~\ref{fig._dissipation_k2dt} in this new model (Fig.~\ref{fig._dissipation_Q_prime}).
\\~\\All the remarks made concerning Fig.~\ref{fig._dissipation_k2dt} can be applied to Fig.~\ref{fig._dissipation_Q_prime}. As accomplished previously, we can constrain the parameter controlling the planetary tidal dissipation. With our grid, we found that  $Q'_{\rm{p}}$ should be higher than $4.5^{+5.8}_{-1.5} \times 10^5$ to reproduce the observed eccentricity. Equivalently, a transition from a regime where the past tidal dissipation is dominated by planetary equilibrium tides towards a dissipation dominated by stellar equilibrium tides appears around $Q'_{\rm{p}}=5.5\times10^6$. 
\\~\\The minimum value of $Q'_{\rm{p}}=4.5^{+5.8}_{-1.5} \times 10^5$ found in this article is compatible with the common value of the dissipation factor for Jupiter found in literature ($Q'_{\rm{p}}=10^6$) \citep{Goldreich1963, Goldreich1966, Leconte2010,Patra2017}. We can mention that the value found is, however, not compatible with the tidal dissipation of the hot Jupiter HAT-P-13b \citep{Batygin2009, Buhler2016}, which is lower than $Q'_{\rm{p}}=3\times10^5$,  found by \cite{Batygin2009} who neglected the impact of the stellar tidal dissipation. The results obtained here can also be compared to the empirical law to find the $Q'_{\rm{p}}$ provided in the work of \cite{Hansen2010,Hansen2012}. When using this law by providing the properties of our system, we found a value for the dissipation factor $Q'_{\rm{p}}=3.2\times10^7$ that is consistent with our findings.
\\~\\To illustrate the evolution of the Kepler-91 system in the different regimes of dissipation, in Fig.~\ref{fig._evolution} we display  the evolution of the semi-major axis and eccentricity of the planet with various values of the tidal dissipation factor  $k_{\rm{2, p}}\Delta t_{\rm{p}}$ used in Fig.~\ref{fig._dissipation_k2dt} and a planet mass of M$_{\rm{pl}}=0.88$ M$_{\rm{Jup}}$.
Figure~\ref{fig._evolution} highlights the significant difference between the regime where past dissipation was dominated by stellar tides with the black curve corresponding to  $k_{\rm{2, p}}\Delta t_{\rm{p}}=0.01$ s and the regime where past tides were dominated by planetary tides, where $k_{\rm{2, p}}\Delta t_{\rm{p}}=0.1$ s and $0.4$ s. In the regime dominated by stellar tides, the dissipation during the stellar MS is slow and the damping of the eccentricity almost null, even at high eccentricity.  During the post-MS, the migration and damping eccentricity become particularly efficient at high eccentricity;  this dissipation is attributed to stellar tides and is  thus  particularly dependent on the stellar parameter and the stellar model. 
\\The opposite happens in the regime where past dissipation was dominated by planetary tides as illustrated in Fig.~\ref{fig._evolution}. The damping of eccentricity and migration is so efficient during the stellar MS due to planetary tides that most of the eccentricity is dissipated before the end of it.  In such scenarios, stellar tides only have a limited impact on the evolution of the system during the stellar post-MS.  Therefore,  in this regime the stellar parameters only have a small impact on the final evolution of the system. 

\subsection{Evolution of the dissipation of the stellar equilibrium tides}\label{subsec_evolution_dissipation_star}

\begin{figure}[h]
\centering
\includegraphics[width=\hsize]{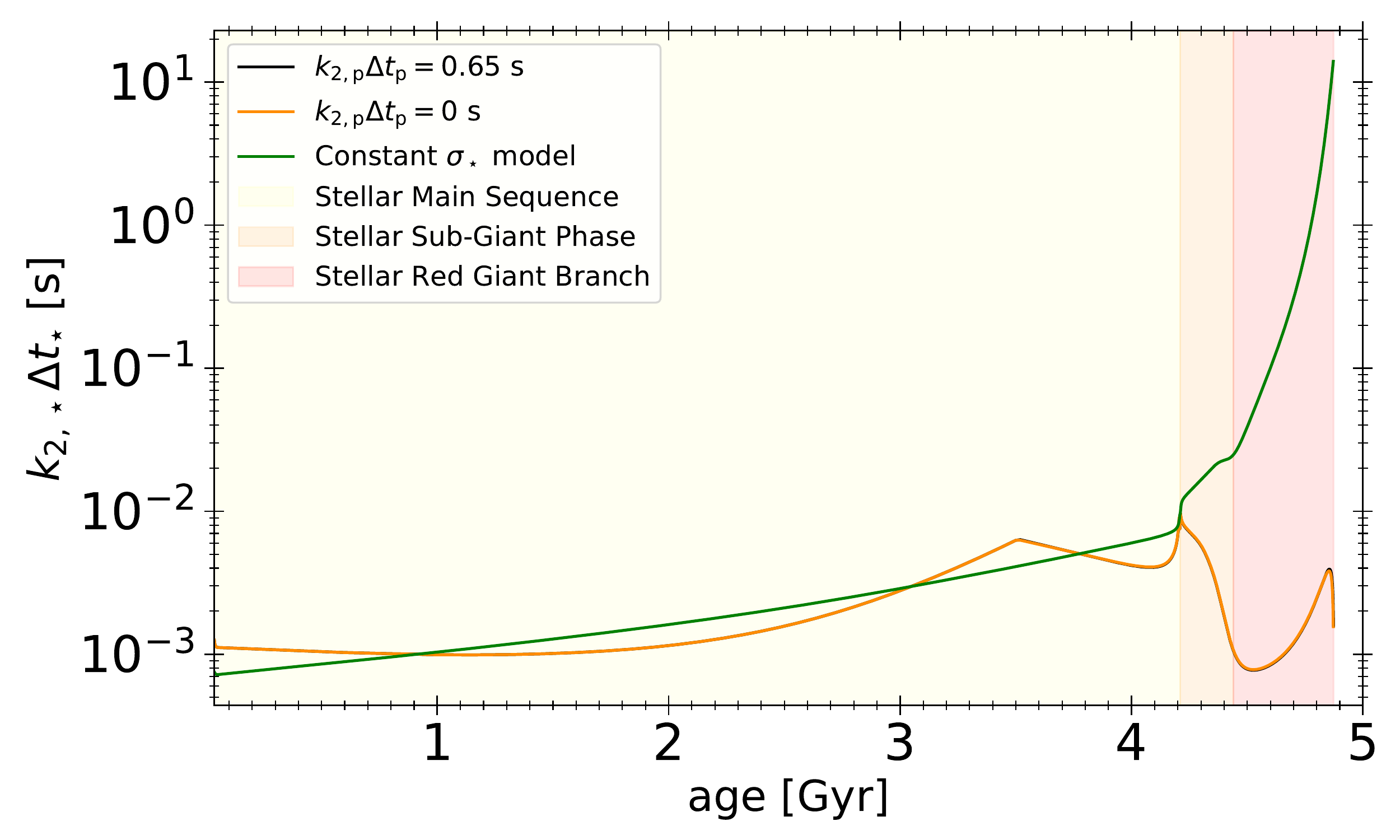}
\caption{Illustration of the evolution of $k_{\rm{2, \star}}\Delta t_\star$ as function of time for the different models considered. In orange our stellar model combined to the upper boundary of the planetary tidal dissipation found previously $k_{\rm{2, p}}\Delta t_{\rm{p}}=0.65$. In black our stellar model is combined to the lower value of the planetary dissipation $k_{\rm{2, p}}\Delta t_{\rm{p}}=0$. We can notice that the black and orange line are close to be superimposed. In green is represented the model with $\sigma_\star=$ const, as done in \cite{Bolmont2016}, the constant was calibrated at value in the MS ($\sigma_\star= 5.536 \times 10^{-66}$ g$^{-1}$cm$^{-2}$s$^{-1}$ at $t=0.83$ Gyr). Finally the shaded areas correspond to the different evolutionary phases of the star, in yellow the MS, in orange, the sub-Giant phase, and in red the RGB phase.}
\label{fig._star_k2Dt}
\end{figure}

In this section, we investigate the tidal dissipation of Kepler-91 starting from the stellar model and the previously derived results.  Usually, the dissipation of the stellar equilibrium tides is expressed with the same formalism as the planetary equilibrium tides, namely the CTL or CPL models. Each of these models are associated with a free parameter controlling the tidal dissipation, $k_{\rm{2, \star}}\Delta t_\star$ for the CTL model and $Q^{\prime}_\star$ for the CPL model.  In our modelling, the study of the dissipation of stellar equilibrium tides does not rely on a free paramenter; on the contrary, the dissipation is deduced from dedicated stellar evolutionary models. We link the stellar tidal dissipation derived in our modelling to the classical dissipation models presented before.
\cite{Bolmont2016} introduced a tidal dissipation factor, $\sigma_\star$, similar to a $k_{\rm{2, \star}}\Delta t_\star$, exclusively relying on the structure and rotation of the star. Following the approach of \cite{Rao2018}, to derive $\sigma_\star$ we can start from Eq. 12 of \cite{Bolmont2016}:
\begin{multline}
\left(\frac{1}{a}\frac{\rm{d} a}{\rm{d} t} \right)_{\rm{eq.tides},\star}  = \\ 9\sigma_\star M_{\star}R_\star q(1+q)\left(\frac{R_\star}{a}\right)^8 \left[ N_{a2}(e)\frac{\Omega_\star}{n} - N_{a1}(e) \right], 
\end{multline}
and we identify the tidal dissipation factor $\sigma_\star$ using the equivalent equation in our model, Eq.~\ref{eq._a_dot_star}; $\sigma_\star$ can then be expressed as 
\begin{equation}\label{eq._sigma_star}
\sigma_\star=\frac{1}{9M_\star R_\star} \frac{f}{\tau}\frac{M_{\rm{env}}}{M_\star}.
\end{equation}
In our expression of the dissipation factor,$\sigma_\star$, the tidal dissipation of the star depends on the orbital period of the planet trough the quantity $f_{\rm{orb}}$ defined in Eq.~\ref{eq._f}. This factor helps to account for the only convective cells that participate in the dissipation process in the regime of fast tides, and it depends on the position of the planetary companion at each evolutionary time step.  Thanks to the constraint on the tides dissipated in the planet, its position is known at each time-step, and consequently it is also possible to compute the $f_{\rm{orb}}$ factor.
\\The tidal dissipation of the star, under the form of $\sigma_\star$, can be linked to $k_{\rm{2, \star}}\Delta t_\star$ with the expression given in \cite{Bolmont2016}: 
\begin{equation}\label{eq._sigma_conversion}
\sigma_\star=k_{\rm{2, \star}}\Delta t_\star\frac{2G}{3R_\star^5},
\end{equation}
where $G$ is the universal gravitational constant. 
\\The constraints on the planetary tidal dissipation given in the previous section are used to constrain the factor $f_{\rm{orb}}$ to determine a range of values of $\sigma_\star,$ and then of $k_{\rm{2, \star}}\Delta t_\star$ using both Eqs.~\ref{eq._sigma_star} and \ref{eq._sigma_conversion}. 
\\Finally, in the models used by \cite{Bolmont2016}, $\sigma_\star$ is assumed constant through the evolution of the system for the stellar equilibrium tides. We also investigated the validity of this approximation with a similar model by setting $\sigma_\star$ to a constant value calibrated at the MS of Kepler-91.
\\The result from our modelling is illustrated in Fig.\ref{fig._star_k2Dt}, where we represent the evolution of $k_{\rm{2, \star}}\Delta t_\star$ as a function of time for the different models considered. Using Eq.~\ref{eq._Q_prime}, we did the same study converting $k_{\rm{2, \star}}\Delta t_\star$ to $Q^{\prime}_\star$; the results are presented in Fig.~\ref{fig._star_q_prime}.
\\~\\Figure~\ref{fig._star_k2Dt} shows that a difference in tidal planetary dissipation and thus on the position of the planet for a given system only has a minor impact on the dissipation factor of the stellar equilibrium tides. We can mention, however, that in Fig.~\ref{fig._star_k2Dt}, the position of the planet dominates the variation of $k_{\rm{2, \star}}\Delta t_\star$ during the inward migration of the planet at the end of the stellar RGB phase.
\\~\\As shown by Fig.~\ref{fig._star_k2Dt}, the dissipation model with a constant $\sigma_\star$ used by \cite{Bolmont2016} is a fair approximation during the stellar MS.  However, such an approximation is inadequate during the PMS and post-MS phases, where changes of the tidal dissipation by orders of magnitude are obtained as a result of the evolution of the internal stellar structure \citep{Rao2018}. This illustrates the need to correctly follow the internal structure of the star during these evolutionary phases, which can either be done by computing full models of the star as in the present study or in our previous works \citep{Pezzotti2021,Pezzotti2021b, Pezzotti2022, Betrisey2022}, or by using pre-calculated grids of stellar models that provide the needed quantities $M_{\rm{env}}$ and $R_{\rm{env,}}$ as done for instance by \cite{Bolmont2016}.

\section{Conclusion}\label{sec_conclusion}


In this study, we carried out a detailed analysis of the properties of Kepler-91, a well-known RGB exoplanet host star observed by \textit{Kepler} \citep{Lillo-Box2014b, Lillo-Box2014}. Our goal was to couple the modelled evolution of the close-in planet to the evolution of the star to study and constrain the dissipation of the planetary equilibrium tides.
\\~\\The strategy of our study is based on a detailed modelling of the entire system. First, the star was modelled independently with the Code Liégeois d'Evolution Stellaire \citep[CLES,][]{Scuflaire2008a} and the Liège OScillation Code \citep[LOSC,][]{Scuflaire2008b}, taking particular care to reproduce the observed classical constrains and asteroseismic quantities. Then, the evolution of the star was coupled with the evolution of the planet by our orbital evolution code \citep{Privitera2016I, Privitera2016II, Meynet2017, Rao2018}. Our orbital evolution code was modified in order to account for the evolution of the eccentricity of the planet aiming at precisely describing the orbital evolution of Kepler 91b in the past and future. In particular, the tidal dissipation of both the star and the planet were taken into account to study the evolution of the planet eccentricity and orbital period.
\\In the modelling of the planetary equilibrium tides, the tidal dissipation is proportional to a free parameter, a dissipation factor, and either $k_{\rm{2, p}}\Delta t_{\rm{p}}  $ \citep{Darwin1879} or $Q'_{\rm{p}}$ \citep{ Goldreich1963, Goldreich1966} depending on the assumptions made. The principle of our study was to determine the maximum eccentricity reachable by the planet, reproducing all the observational constraints and given a tidal dissipation factor. 
\\~\\With this methodology, we found that $k_{\rm{2, p}}\Delta t_{\rm{p}}$ should be lower than $0.4\pm0.25$s to be able to reproduce the observed eccentricity of the planet. We also found that $Q'_{\rm{p}}$ should be greater than $4.5^{+5.8}_{-1.5} \times 10^5$. We then looked at the dissipation of the stellar equilibrium tide, which was computed based on the physical description of \cite{Zahn1966, Zahn1977, Zahn1978, Zahn1989, Villaver2009, Villaver2014} and \cite{Villaver2014} with the stellar structure given by our stellar modelling. The tidal dissipation from our modelling was compared to a more classical formalism such as the $\sigma_\star$ formalism \citep{Bolmont2016}. We confirm previous results of \cite{Rao2018} that a constant $\sigma_\star$  is a correct approximation during the stellar MS, but not during the PMS and post-MS phases, where the evolution of the internal stellar structure needs to be followed in detail. 
\\~\\More than putting a constraint on the planetary dissipation of the planet equilibrium tides, our study shows that hot Jupiters have tidal dissipation coefficients in the same range of values as colder, Jupiter-like planets. This result can be important for the theoretical study of multi-planetary systems as tidal interactions between planets play a key role in the understanding of their future evolution and stability. Moreover, our results show that planetary structural modifications caused by the extended atmosphere of hot Jupiters do not significantly increase the tidal dissipation factors.
\\~\\The main limits to the present method come from the overall precision of the characterisation of the system, and in particular the planetary properties.
In the future, the same type of study could be applied to similar targets characterised with more accuracy. The ideal target should be orbiting an evolved star with observed oscillation modes to conduct a similar precise stellar modelling; thus, we would favour a \textit{Kepler} \citep{KEPLER2010} or TESS \citep{TESS2014, TESS2015} target. The studied planet needs to be in close orbit with respect to the star and have a high eccentricity; in this aspect, the target presented in \cite{Grunblatt2018, Grunblatt2022} could be a good choice. Finally, the characterisation of the planet, and in particular a precise determination of the planetary mass, is necessary. The use of other constraints such as the planet period variation or precession variation can also provide important information about the tidal dissipation of the planet.
\begin{acknowledgements} The authors are thanking the anonymous referee for his constructive comments.
L.F was supported by the Fonds de la Recherche Scientifique$–$ FNRS as a Reaserch Fellow.
C.P. acknowledges support by the Swiss National Science Foundation (project number 200020-205154).
P.E. have received funding from the European Research Council (ERC) under the European Union's Horizon 2020 research and innovation programme (grant agreement No 833925, project STAREX).
G.B. acknowledges funding from the SNF AMBIZIONE grant No. 185805 (Seismic inversions and modelling of transport processes in stars).
This article used an adapted version of InversionKit, a software developed within the HELAS and SPACEINN networks, funded by the European Commissions's Sixth and Seventh Framework Programmes. This work has been carried out within the framework of the NCCR PlanetS supported by the Swiss National Science Foundation. The computations were performed at University of Geneva on the Baobab and Yggdrasil clusters. This research has made use of NASA's Astrophysics Data System.
\end{acknowledgements}

\bibliographystyle{aa}
\bibliography{biblio_KEP-91}

\begin{appendix}
\section{Seismic data}
\begin{table}[h]
\caption{Seismic data obtained by \cite{Lillo-Box2014} and used in this article. }
\begin{tabular}{ccc}\label{table_data_sismo}
Degree $\ell$        & Frequency $\nu\ [\mu \rm{Hz}]$  \\ \hline
2                    & $76.835 \pm 0.022$                                     \\
0                    & $78.160 \pm 0.031$                                     \\
2                    & $85.924 \pm 0.014$                                     \\
0                    & $87.156 \pm 0.019$                                     \\
2                    & $95.004 \pm 0.020$                                      \\
0                    & $96.289 \pm 0.016$                                      \\
2                    & $101.929 \pm 0.010$                                     \\
0                    & $105.792 \pm 0.012$                                      \\
2                    & $114.018 \pm 0.018$                                      \\
0                    & $115.159 \pm 0.011$                                      \\
2                    & $123.468 \pm 0.028$                                      \\
0                    & $124.663 \pm 0.024$                                      \\
2                    & $133.215 \pm 0.030$                                      \\
0                    & $134.326 \pm 0.046$                                      \\
\multicolumn{1}{l}{} & \multicolumn{1}{l}{}                                   \\
1                    & $73.510 \pm 0.035$                                     \\
1                    & $82.271 \pm 0.013$                                     \\
1                    & $82.720 \pm 0.023$                        \\
1                    & $83.115 \pm 0.020$                                     \\
1                    & $91.514 \pm 0.012$                     \\
1                    & $91.913 \pm 0.012$                   \\
1                    & $92.386 \pm 0.017$                    \\
1                    & $92.958 \pm 0.025$                    \\
1                    & $99.843 \pm 0.009$                 \\
1                    & $101.146 \pm 0.028$                    \\
1                    & $101.345 \pm 0.013$                  \\
1                    & $101.929 \pm 0.007$                   \\
1                    & $110.459\pm 0.043$                    \\
1                    & $110.855 \pm 0.019$                    \\
1                    & $110.995 \pm 0.035$                    \\
1                    & $111.574 \pm 0.011$                    \\
1                    & $119.546 \pm 0.021$                    \\
1                    & $120.198 \pm 0.012$                    \\
1                    & $121.002 \pm 0.013$                    \\
1                    & $129.008 \pm 0.017$                    \\
1                    & $129.783 \pm 0.027$                    \\
1                    & $138.483 \pm 0.040$                    \\
1                    & $139.520 \pm 0.023$                    \\

\multicolumn{2}{l}{}                                                               \\
\multicolumn{2}{l}{Large separation: $\Delta\nu=9.39 \pm 0.22\ \mu \rm{Hz}$}              \\
\multicolumn{2}{l}{Frequency of max. power: $\nu_{max}=108.9 \pm 3.0\ \mu \rm{Hz}$}      
\end{tabular}
\end{table}
\FloatBarrier
\section{Additional tables}

\begin{table*}[h]
\caption{Initial and final stellar,  orbital and planetary parameters obtained for each  model made in our parameter space exploration to constrain the tidal dissipation factor. The column with the final eccentricity corresponds to the maximum eccentricity reachable by the planet at the current age of the system.  This eccentricity corresponds to the maximum eccentricity presented in Fig.~\ref{fig._dissipation_k2dt}.}.\label{table_dissipation_constraint}
\resizebox{\hsize}{!}{\begin{tabular}{llllllllllll}
\hline\hline
\multicolumn{4}{c}{initial system properties}                                                                                                                                       &  & \multicolumn{2}{c}{free parameters}                                                     &  & \multicolumn{4}{c}{final system properties}                                                                                                                          \\ \cline{1-4} \cline{6-7} \cline{9-12} 
                                                       &                                  &                         &                                                               &  &                                                              &                                  &  &                                    &                         &                                                        &                                              \\
\multicolumn{1}{c}{M$_{\rm{pl}}$ {[}M$_{\rm{Jup}}${]}} & \multicolumn{1}{c}{$a$ {[}AU{]}} & \multicolumn{1}{c}{$e$} & \multicolumn{1}{c}{$\Omega_\star$ {[}$\Omega_{\rm{crit}}${]}} &  & \multicolumn{1}{c}{$k_{\rm{2, p}}\Delta t_{\rm{p}}$ {[}s{]}} & \multicolumn{1}{c}{$f_{\rm{r}}$} &  & \multicolumn{1}{c}{Period {[}d{]}} & \multicolumn{1}{c}{$e$} & \multicolumn{1}{c}{R$_{\rm{pl}}$ {[}R$_{\rm{Jup}}${]}} & \multicolumn{1}{c}{$\Omega_\star$ {[}nHz{]}} \\
                                                       &                                  &                         &                                                               &  &                                                              &                                  &  &                                    &                         &                                                        &                                              \\ \hline\hline
0.55                                                   & 0.4000                           & 0.90                    & 0.4                                                          &  & 0.01                                                         & 1.345                            &  & 6.24668                            & 0.433                   & 1.383                                                  & 263                                          \\
0.55                                                   & 0.3199                           & 0.90                    & 0.4                                                          &  & 0.1                                                          & 1.345                             &  & 6.24668                            & 0.117                   & 1.383                                                  & 261                                          \\
0.55                                                   & 0.3239                           & 0.90                    & 0.4                                                          &  & 0.25                                                         & 1.345                             &  & 6.24668                            & 0.037                   & 1.383                                                  & 257                                          \\
0.55                                                   & 0.3240                           & 0.90                    & 0.4                                                          &  & 0.5                                                          & 1.345                             &  & 6.24669                            & $1\times10^{-5}$        & 1.384                                                  & 255                                          \\
0.55                                                   & 0.3240                           & 0.90                    & 0.4                                                          &  & 0.75                                                         & 1.345                             &  & 6.24669                            & $7\times10^{-8}$        & 1.384                                                  & 255                                          \\
0.55                                                   & 0.3240                           & 0.90                    & 0.4                                                          &  & 1                                                            & 1.345                             &  & 6.24669                            & $3\times10^{-10}$       & 1.384                                                  & 255                                          \\
                                                       &                                  &                         &                                                               &  &                                                              &                                  &  &                                    &                         &                                                        &                                              \\
0.88                                                   & 0.4229                           & 0.90                    & 0.4                                                          &  & 0.01                                                         & 1.345                            &  & 6.24668                            & 0.402                   & 1.384                                                  & 265                                          \\
0.88                                                   & 0.3201                           & 0.90                    & 0.4                                                          &  & 0.1                                                          & 1.345                            &  & 6.24668                            & 0.298                   & 1.384                                                  & 262                                          \\
0.88                                                   & 0.3233                           & 0.90                    & 0.4                                                          &  & 0.25                                                         & 1.345                            &  & 6.24668                            & 0.141                   & 1.384                                                  & 261                                         \\
0.88                                                   & 0.3271                           & 0.90                    & 0.4                                                          &  & 0.5                                                          & 1.345                            &  & 6.24668                            & 0.033                   & 1.384                                                  & 257                                          \\
0.88                                                   & 0.3271                           & 0.90                    & 0.4                                                          &  & 0.75                                                         & 1.345                            &  & 6.24668                            & 0.001                   & 1.384                                                  & 257                                          \\
0.88                                                   & 0.3271                           & 0.90                    & 0.4                                                          &  & 1                                                            & 1.345                            &  & 6.24668                            & 0.001                  & 1.384                                                  & 257                                          \\
                                                       &                                  &                         &                                                               &  &                                                              &                                  &  &                                    &                         &                                                        &                                              \\
1.05                                                   & 0.4358                           & 0.90                    & 0.4                                                          &  & 0.01                                                         & 1.345                            &  & 6.24667                            & 0.385                   & 1.383                                                  & 268                                          \\
1.05                                                   & 0.3294                           & 0.90                    & 0.4                                                          &  & 0.1                                                          & 1.345                            &  & 6.24667                            & 0.368                   & 1.383                                                  & 267                                          \\
1.05                                                   & 0.3309                           & 0.90                    & 0.4                                                          &  & 0.25                                                         & 1.345                            &  & 6.24668                            & 0.207                   & 1.383                                                  & 264                                          \\
1.05                                                   & 0.3313                           & 0.90                    & 0.4                                                          &  & 0.5                                                          & 1.345                            &  & 6.24669                            & 0.085                   & 1.384                                                  & 261                                          \\
1.05                                                   & 0.3317                           & 0.90                    & 0.4                                                          &  & 0.75                                                         & 1.345                            &  & 6.24669                            & 0.029                   & 1.384                                                  & 261                                          \\
1.05                                                   & 0.3317                           & 0.90                    & 0.4                                                          &  & 1                                                            & 1.345                            &  & 6.24669                            & 0.009                   & 1.384                                                  & 261                                          \\ \hline\hline
\end{tabular}
}
\end{table*}
\FloatBarrier

\section{Constraints on the tidal dissipation with the CPL model}
\begin{figure}[h]
\centering
\includegraphics[width=\hsize]{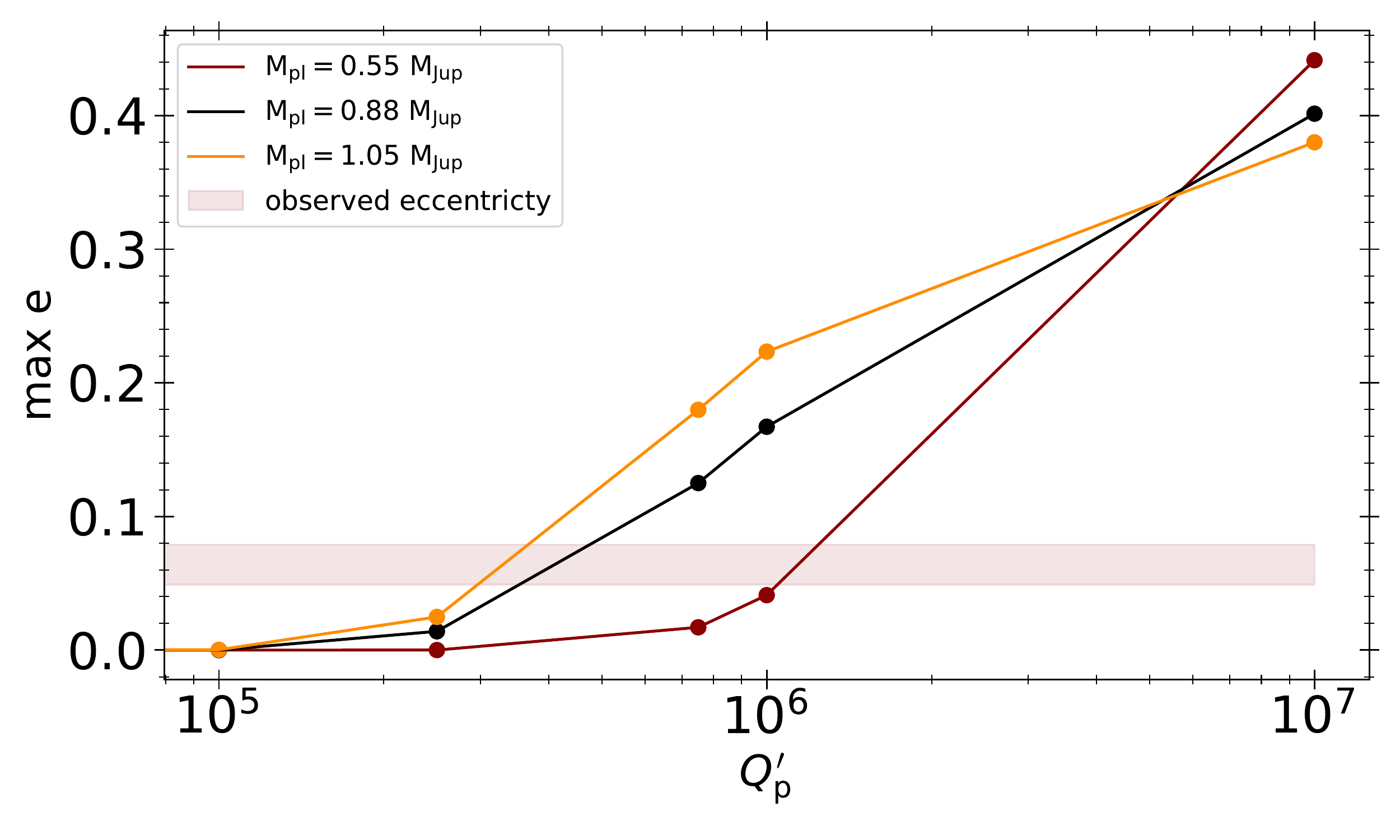}
\caption{Results from the grid of evolutionnary tracks where the maximum eccentricity of Kepler 91b is given for some values of the tidal dissipation parameter $k_{\rm{2, p}}\Delta t_{\rm{p}}$. These results are compared to the observed eccentricity for the three different planetary masses considered.}
\label{fig._dissipation_Q_prime}
\end{figure}
\FloatBarrier

\section{Additional figures}

\begin{figure}[h]
\centering
\includegraphics[width=\hsize]{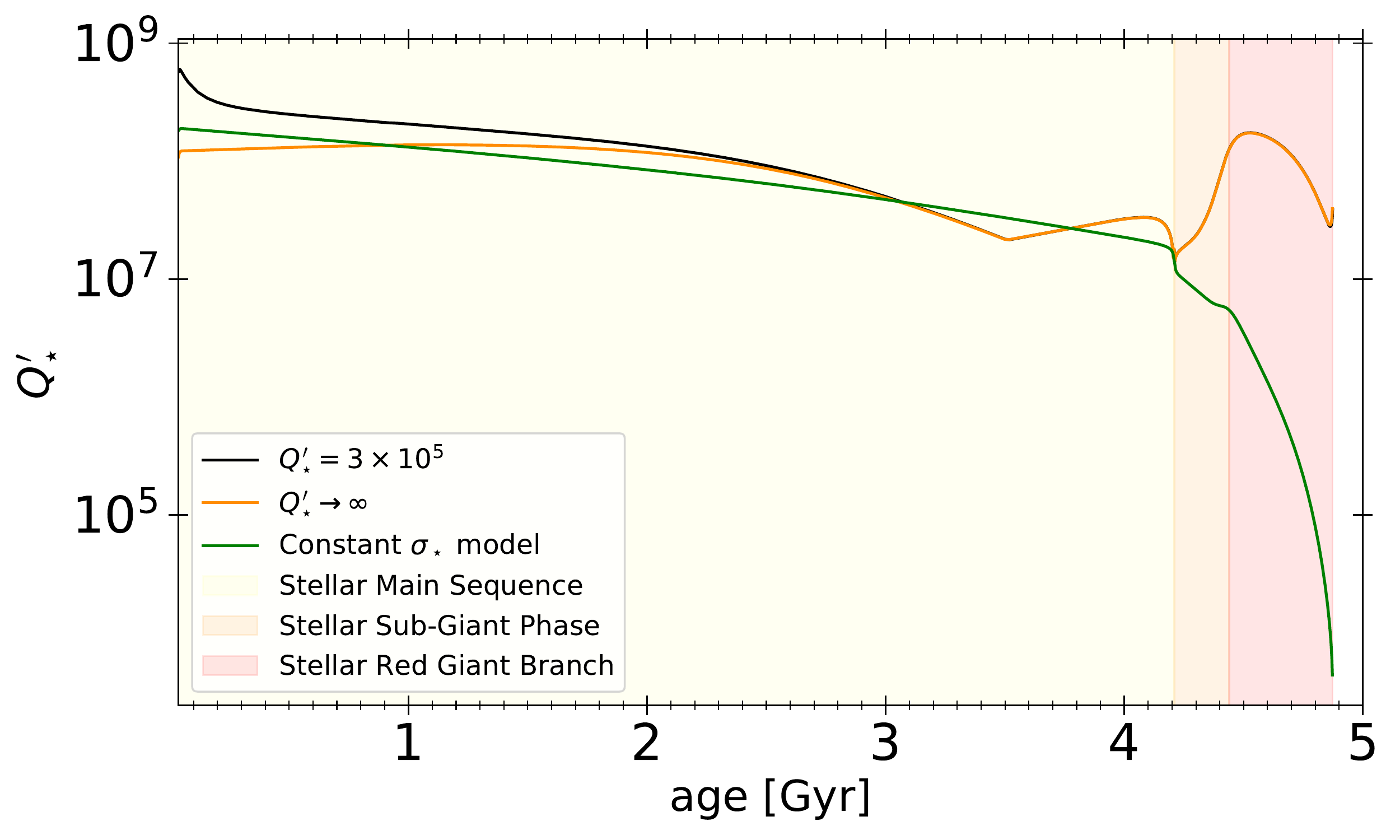}
\caption{Illustration of the evolution of $Q^\prime_\star$ as function of time for the different models considered. In orange our stellar model combined to the upper boundary of the planetary tidal dissipation found previously $Q^\prime_\star=3\times10^5$. In black our stellar model is combined to the lower value of the planetary dissipation $Q^\prime_\star\rightarrow\infty$. We can notice that the black and orange line are close to be superimposed. In green is represented the model with $\sigma_\star=$ const, as done in \cite{Bolmont2016}, the constant was calibrated at value in the MS ($\sigma_\star= 5.536 \times 10^{-66}$ at $t=0.83$ Gyr). Finally the shaded areas correspond to the different evolutionary phases of the star, in yellow the MS, in orange, the sub-Giant phase, and in red the RGB phase.}
\label{fig._star_q_prime}
\end{figure}
\FloatBarrier

\end{appendix}

\end{document}